\documentclass[a4paper,11pt]{article}
\usepackage{pos}

\let\OLDthebibliography\thebibliography
\renewcommand\thebibliography[1]{
  \OLDthebibliography{#1}
  \setlength{\parskip}{0pt}
  \setlength{\itemsep}{0pt plus 0.3ex}
}

\usepackage{lineno}

\usepackage{wrapfig}

\title{All-Sky Cosmic-Ray Anisotropy Update at Multiple Energies}

\ShortTitle{All-Sky Cosmic-Ray Anisotropy Update at Multiple Energies}

\author{The IceCube \& HAWC Collaborations\\{\normalsize \normalfont(a complete list of authors can be found at the end of the proceedings)}\\}

\emailAdd{juancarlos@icecube.wisc.edu}
\emailAdd{rykore@wisc.edu}
\emailAdd{paolo.desiati@icecube.wisc.edu}
\emailAdd{fawolf@wisc.edu}

\abstract{
We present preliminary results on an updated full-sky analysis of the cosmic-ray arrival direction distribution with data collected by the High-Altitude Water Cherenkov (HAWC) Observatory and IceCube Neutrino Observatory with complementary field of views covering a large fraction of the sky. This study extends the energy range to higher energies. The HAWC Observatory, located at 19°N has analyzed 8 years of cosmic-ray data over an energy range between 3.0 TeV and 1.0 PeV and confirms an energy-dependent anisotropy in the arrival direction distribution of cosmic rays seen by other experiments. Combined with recently published results from IceCube with 12 years of data, the combined sky maps with 93\% coverage of the sky —between 70°N and 90°S— and the corresponding angular power spectra largely eliminate biases that result from partial sky coverage.

\vspace{4mm}

{\bfseries Corresponding authors:}
Juan Carlos Díaz Vélez$^{1*}$, 
Riya Yogesh Kore$^{1}$,
Paolo Desiati$^{1}$,
Ferris Wolf$^{1}$
\\
{$^{1}$ \itshape Dept. of Physics and Wisconsin IceCube Particle Astrophysics Center, University of Wisconsin{\textemdash}Madison, Madison, WI 53706, USA}\\
[4mm]
$^*$ Presenter
}

\ConferenceLogo{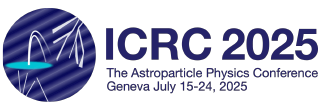}

\FullConference{39th International Cosmic Ray Conference (ICRC2025)\\
 15–24 July 2025\\
Geneva, Switzerland\\}

\vspace{4mm}

\ConferenceLogo{PoS_ICRC2025_logo.pdf}

\FullConference{39th International Cosmic Ray Conference (ICRC2025)\\
 15–24 July 2025\\
Geneva, Switzerland\\}

\begin{document}

\maketitle

\section{Introduction}
Over the last few decades, several ground-based experiments in both the northern and southern hemispheres have observed variations at the level of $10^{-3}$ in the arrival direction distribution of cosmic rays with energies between 1~TeV to several PeVs with high statistical accuracy (see ~\citep{Abbasi_2025} and references therein).
The limited integrated field of view (FoV) of the sky in all of these individual measurements restricts our ability to characterize the anisotropy in terms of its spherical harmonic components,  
as large correlations between terms $a_{\ell m}$ bias the interpretation of the cosmic ray distributions~\citep{SOMMERS2001271,Ahlers_2019}. 
In this analysis, we apply the same methods used in~\citep{Ahlers_2019} to combine data from the IceCube~\citep{Aartsen:2016nxy} and HAWC~\citep{ABEYSEKARA2023168253} observatories and produce maps with near full--sky coverage to study arrival direction distributions at different energies.

\section{Data}
The northern hemisphere data used for this analysis were collected by the HAWC gamma-ray Observatory over the span of 8 years from May 2015 to May 2023. The data were split into 11 energy bins. In HAWC, the primary energy was estimated using the logarithm of the number of photomultiplier tubes (PMTs) triggered by each event ($\log_{10}N_\mathrm{hit}$), the corresponding reconstructed shower zenith angle ($\cos{\theta_\mathrm{reco}}$), and the radial distance from the center of the HAWC array to the reconstructed shower core location ($R_\mathrm{core}$). A four--dimensional histogram including the logarithm of the true primary cosmic--ray energy ($\log_{10}E_\mathrm{true}$) is fit with a B-spline function~\citep{WHITEHORN20132214}, which provides a smooth lookup table for the median energy of cosmic rays in a similar way as in~\cite{aartsen2016}.
The IceCube dataset is described in detail in \citep{Abbasi_2025}, though energy cuts have been adjusted to match the data from comparable HAWC maps as will be discussed in section \ref{sec:combined}. Table~\ref{tab:energy_bins} shows the number of events in each of the 11 datasets along with the corresponding median energy and rigidity.

\begin{table}[h]
\footnotesize
\centering
\begin{tabular}{l|c|c|c|c}
\hline
Bin & Rigidity (TV)  &	HAWC Energy (TeV) & IceCube Energy (TeV) & Events \\
\hline
0& 0.6 & 1.8 &  -- & $33.7\times 10^{9}$\\
1& 1.1 & 3.2 &  -- & $58.0\times 10^{9}$\\
2& 2.2 & 5.6 &  -- & $44.2\times 10^{9}$ \\
3& 4.1& 10.0 &  -- & $33.4\times 10^{9}$ \\
4& 7.4  & 17.8  &  10.0   & $3\times 10^{11}$\\
5& 12.5 & 31.6  &  14.8 & $2.9\times 10^{11}$\\ 
6&  23.2 & 56.2  &  30.2  & $1.1\times 10^{11}$ \\
7& 40.0 & 100.0  &   53.4 & $3.2\times 10^{10}$\\
8&  75.5 & 177.8  &  310.5  & $1.9\times 10^9$\\
9& 140.9 & 316.2  &  725.3  & $4.3\times 10^9$\\
10& 280.5 & 562.3  &  1716 & $9.9\times 10^7$\\
\hline
\end{tabular}
\caption{Median cosmic--ray primary energy bins used in this analysis, along with corresponding rigidity according to the GST composition model and number of events for each map.}\label{tab:energy_bins}
\end{table}

\subsection{Surface Pressure and Solar Dipole}
Atmospheric conditions, particularly pressure and temperature, can influence cosmic ray detection rates at observatories like HAWC and IceCube.
The HAWC Observatory is located below the shower maximum ($X_\mathrm{max}$) for energies around 10 TeV. As a result, an increase in surface pressure leads to greater atmospheric overburden, attenuating air showers and reducing the event rate. This anti-correlation 
is quantified with a barometric coefficient, which we have determined experimentally to be $\beta=-0.0086\, \mathrm{hPa}^{-1}$ from local measurements. While many of the pressure variations are due to weather, there is a bi-diurnal variation caused by a tidal effect from solar heating. This variation can introduce a spurious signal in the presence of large time gaps in data taking. The data are weighted to compensate for this effect, leaving minimal residual effects. In contrast, IceCube’s muon rate is positively correlated with stratospheric temperature and exhibits strong seasonal variations due to its polar location. Although these atmospheric effects could modulate the Solar dipole signal, their influence is minor in the case of IceCube (see ~\citep{VERPOEST2024102985} and references therein).

In addition to this spurious signal, the Earth’s revolution around the Sun produces a faint Compton-Getting dipole anisotropy~\cite{Compton_1935} with an excess oriented towards the direction of motion in solar coordinates. The relative rotation of the celestial and solar reference frames over a calendar year causes interference between the two sources of anisotropy. This dipole has a predictable phase and amplitude given by
\begin{equation}
\frac{\delta I}{I} = 
\frac{v(t)}{c}\left[\gamma(E)+2\right]\cos{\xi}\,,
\label{eq:compton_getting}
\end{equation}
where $I$ the cosmic ray intensity, $\gamma(E)$ the cosmic--ray spectral index, $v(t)/c$ the ratio of Earth’s orbital
velocity to the speed of light, and $\xi$ the angle between the cosmic ray particle’s arrival direction and the direction of Earth's motion. As with the pressure variations, we can compensate by weighting each event accordingly. The combined weight, with the pressure correction for event $i$, is given by 
\begin{equation}
 w_{i} = e^{-\beta(p(t_i)- p_0)}\left(1-\frac{v(t_i)}{c}\left[\gamma(E_i)+2\right]\cos{\xi_i}\right)\,,
\label{eq:correction_weight}
\end{equation}
where $p(t_i)$ is the measured pressure at the time $t_i$, $p_0$ is the baseline average pressure over time, $v(t_i)$ corresponds to Earth's orbital speed at time $t_i$ (obtained from the PAL astronomical library~\cite{2013ASPC..475..307J}) and $\gamma(E_i)$ is the spectral index at energy $E_i$ derived from a fit to the measurement of the all-particle cosmic-ray spectrum by HAWC~\cite{Morales-Soto:2023rX}.

\section{Relative Intensity}
The relative intensity in J2000 equatorial
coordinates ($\alpha$, $\delta$) is obtained by binning the sky into pixels of size 0.9$^\circ$ using the 
{\tt HEALPix} library~\citep{Gorski:2005apr}. 
 The residual anisotropy $\delta I$ of the distribution of arrival directions of the cosmic rays is calculated by subtracting a reference map that describes the detector response to an isotropic flux
\begin{equation}
\delta I_j =  \frac{N_j- \langle N_j  \rangle}{\langle N_j  \rangle}.
\label{eq:ri}
\end{equation}
This relative intensity gives the amplitude of deviations in the number of counts $N_j$ from the isotropic
expectation $\langle N_j \rangle$ in each pixel  $j$.

The isotropic expectation $\langle N_j \rangle$ and relative intensity are calculated using a likelihood-based reconstruction method for combined data sets from multiple observatories that have overlapping exposure regions in the sky. The method is described in~\cite{0004-637X-823-1-10} and does not rely on detector simulations, providing an optimal anisotropy reconstruction and the recovery of the large-scale anisotropy projected onto the equatorial plane. 

\section{Statistical Significance}
In order to calculate the statistical significance of anisotropy features in the final reconstructed map, we apply a generalized version (\cite{0004-637X-823-1-10}) of the Li--Ma method in~\citep{Li:1983fv}.
The significance map is calculated using 
\begin{equation}\label{eq:significance}
  S_{ i } = \sqrt{2}\left(-\mu_{ i ,\rm on}+\mu_{ i ,\rm off} + n_ i \log\frac{\mu_{ i ,\rm on}}{\mu_{ i ,\rm off}}\right)^{1/2}\,.
\end{equation}
For each pixel $i$, we define expected {\it on-source} and {\it off-source} event counts from neighbor pixels in a disc of radius $r$ centered on that pixel. 
\begin{figure*}[hbtp]
  \centering
\includegraphics[width=0.45\textwidth]{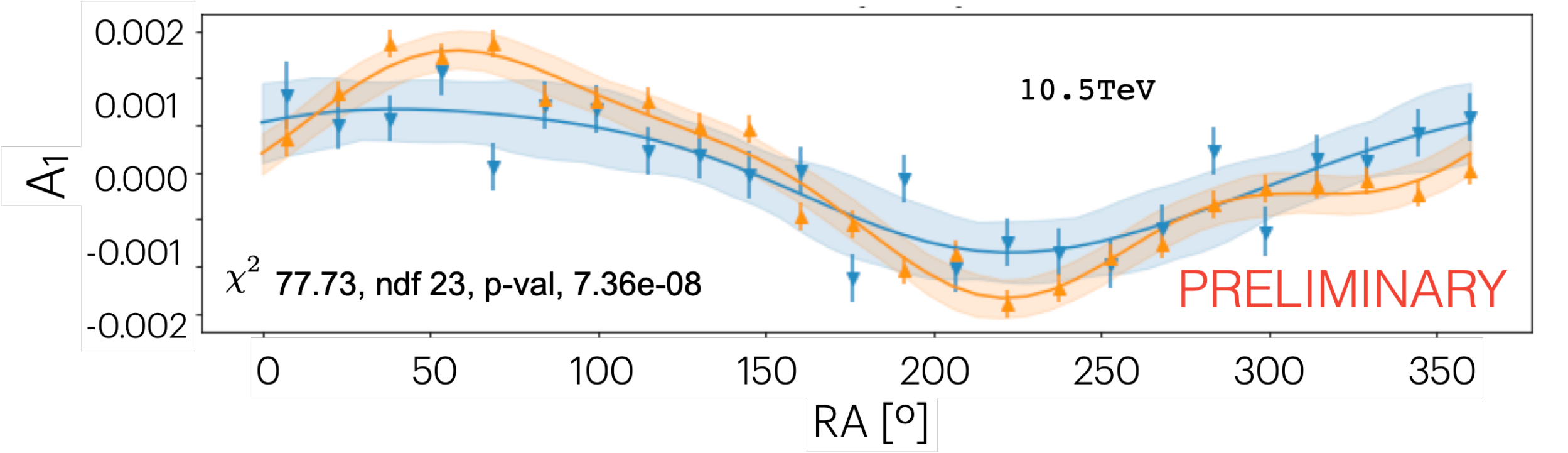}
\includegraphics[width=0.45\textwidth]{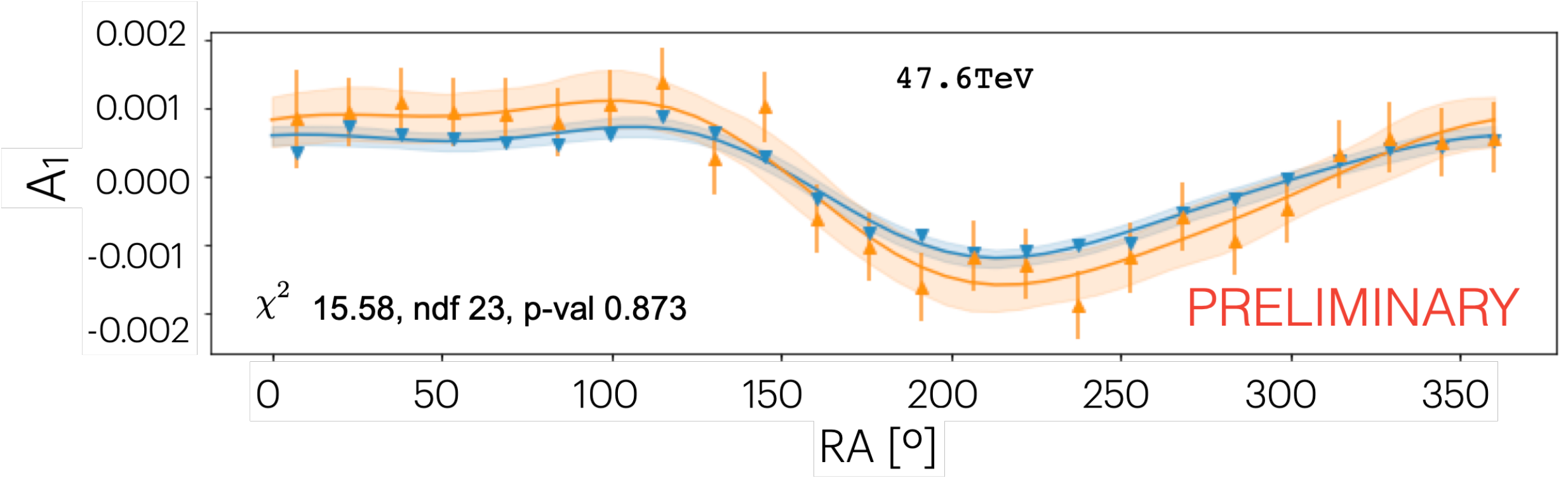}
\includegraphics[width=0.45\textwidth]{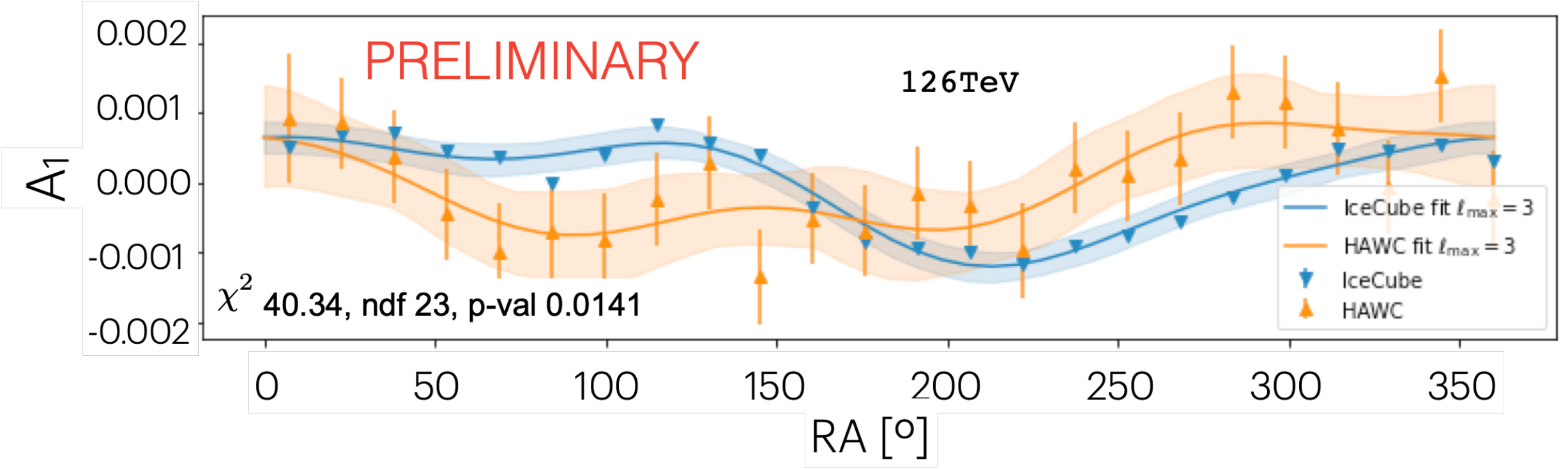}
  \caption{\footnotesize One-dimensional projection of relative intensity in the overlapping field of view of IceCube and HAWC. 
  At 48~TeV energies, there is good statistical agreement between both experiments (top right). However at lower and higher energies (top left and center bottom),
   the two distributions are not statistically compatible.}
  \label{fig:overlapp_orig}
\end{figure*}

\section{Combined Anisotropy with IceCube}\label{sec:combined}
\begin{figure*}[hbtp]
  \centering
  \includegraphics[width=0.56\textwidth]{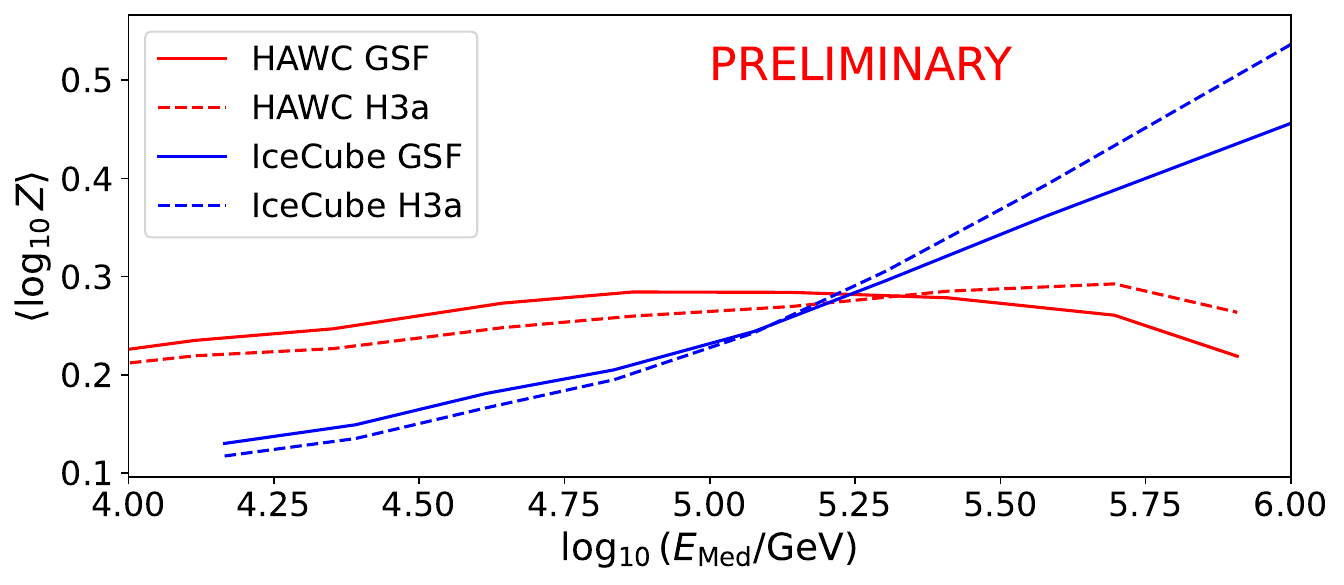}
  \includegraphics[width=0.37\textwidth]{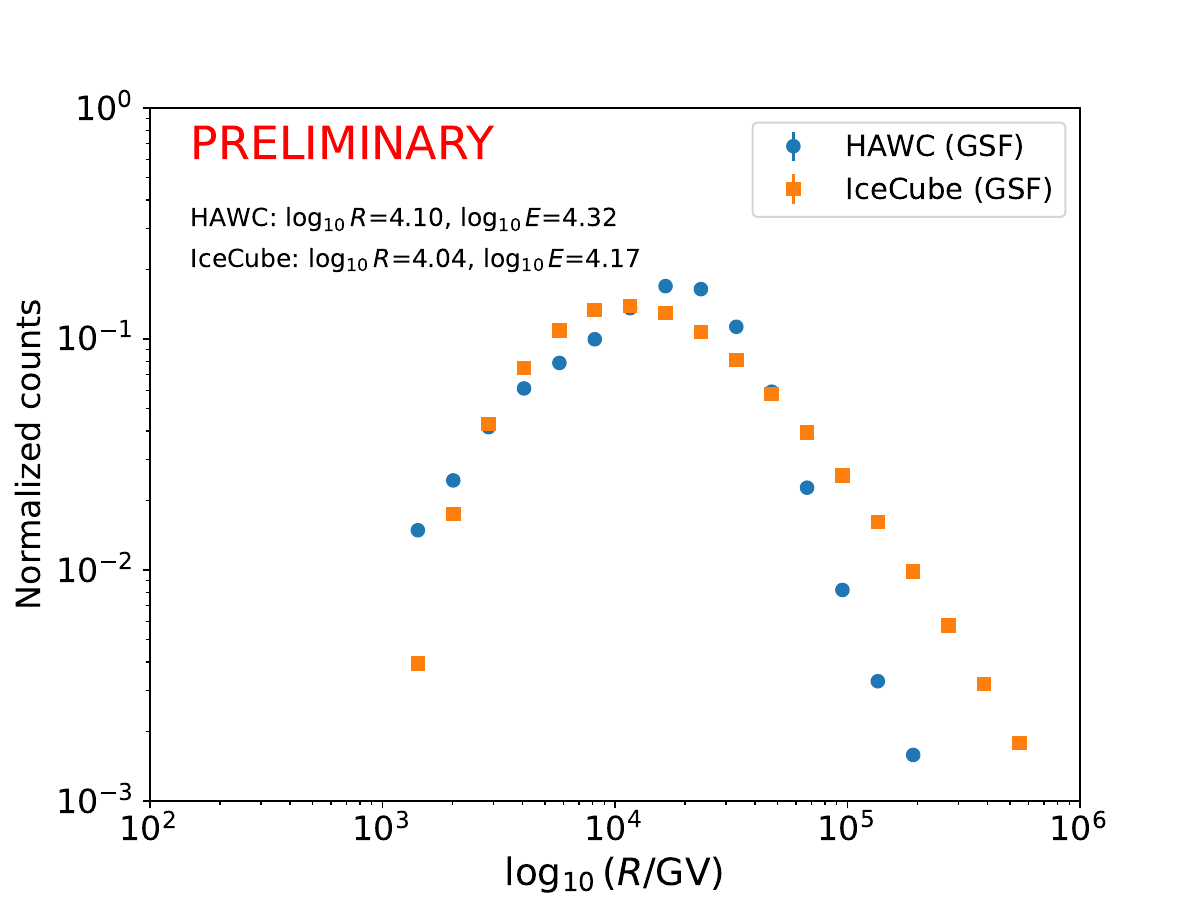}
  \caption{\footnotesize Left: mean logarithm of the particle charge Z for cosmic rays detected by IceCube (blue) and HAWC (red) assuming the Gaisser H3a~\citep{Gaisser:2012zz} (dashed) and GSF~\citep{Dembinski:2017N7} (solid) composition models. IceCube data is dominated by protons and light elements at lower energies than HAWC, but becomes heavier above about 200~TeV.
  Right: 
  Assuming a rigidity--dependent angular distribution of cosmic rays, we find the most compatible energy bins (top) by sliding an IceCube energy window of size $\Delta\log_{10}E = 0.25$ and minimizing the KS test value between rigidity distributions based on Monte Carlo simulations.}
  \label{fig:mc_logz}
\end{figure*}

The overlapping FoV between the two observatories serves as a calibration region. Fig.~\ref{fig:overlapp_orig} shows a one--dimensional projection of the relative intensity as a function of R.A. At 10~TeV energy, the distributions are qualitatively different but are statistically compatible. At 48~TeV energies, there is good agreement between both experiments. At 126~TeV, the distributions are not statistically compatible. 
According the simulations, the cosmic-ray energy distribution in each bin should be comparable, however, the mass composition in each of the samples shown in Fig.~\ref{fig:mc_logz}, indicates that the two detectors have very different sensitivities to each of the mass groups. IceCube data appears to be dominated by protons and light elements at lower energies than HAWC but becomes heavier at $\sim150$~TeV.

\begin{figure*}[hbtp]
  \centering
  \includegraphics[width=0.30\textwidth]{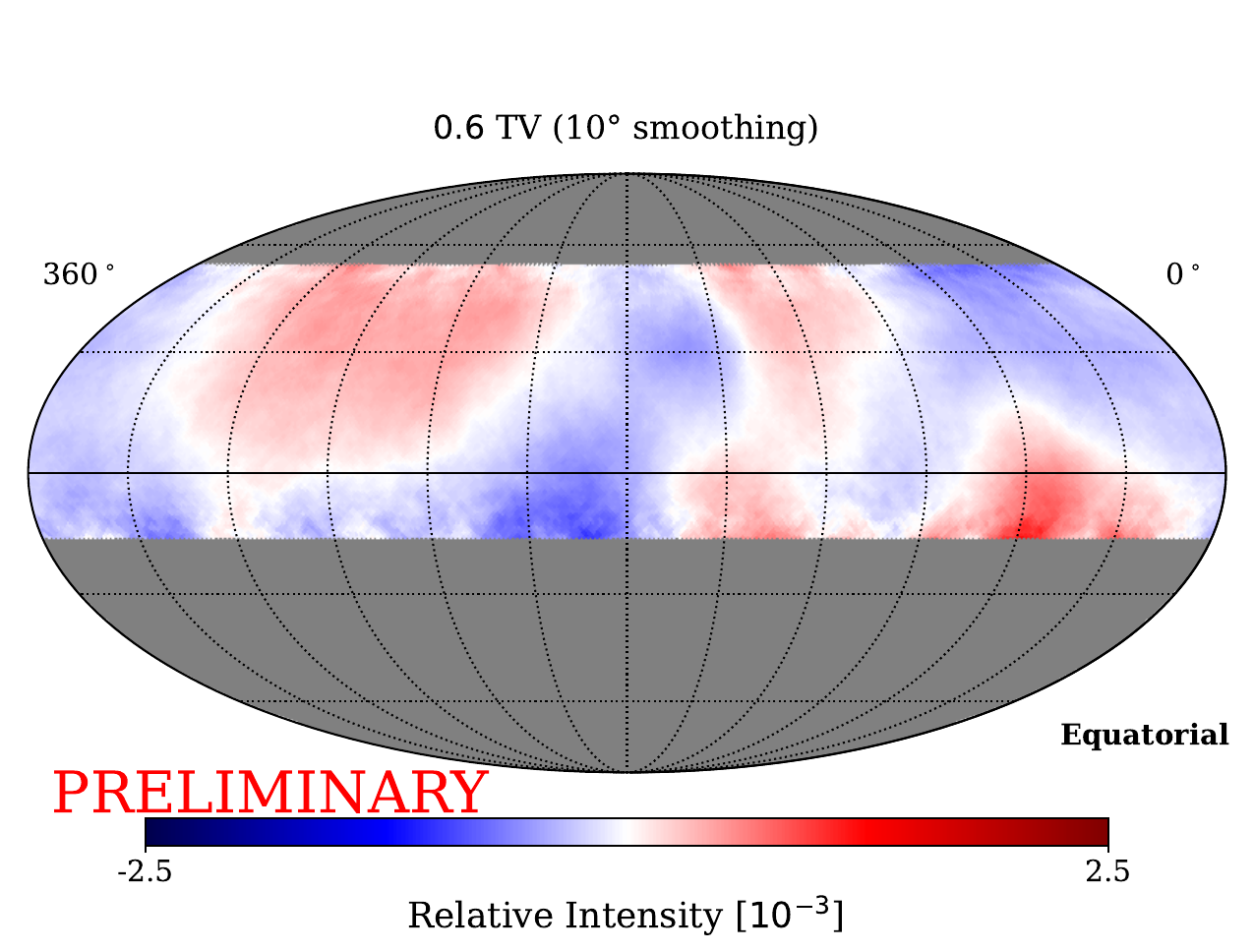}
  \includegraphics[width=0.30\textwidth]{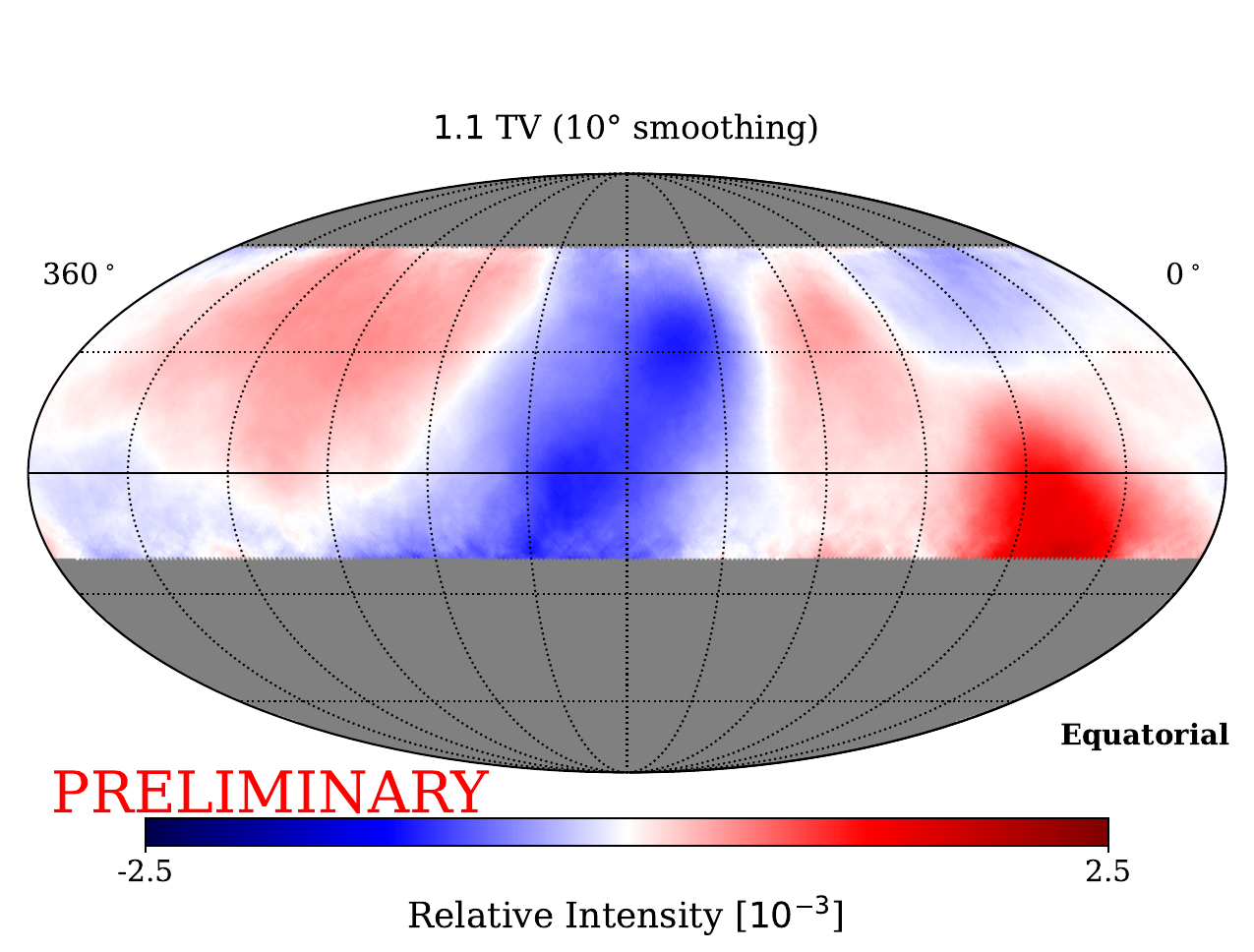}
  \includegraphics[width=0.30\textwidth]{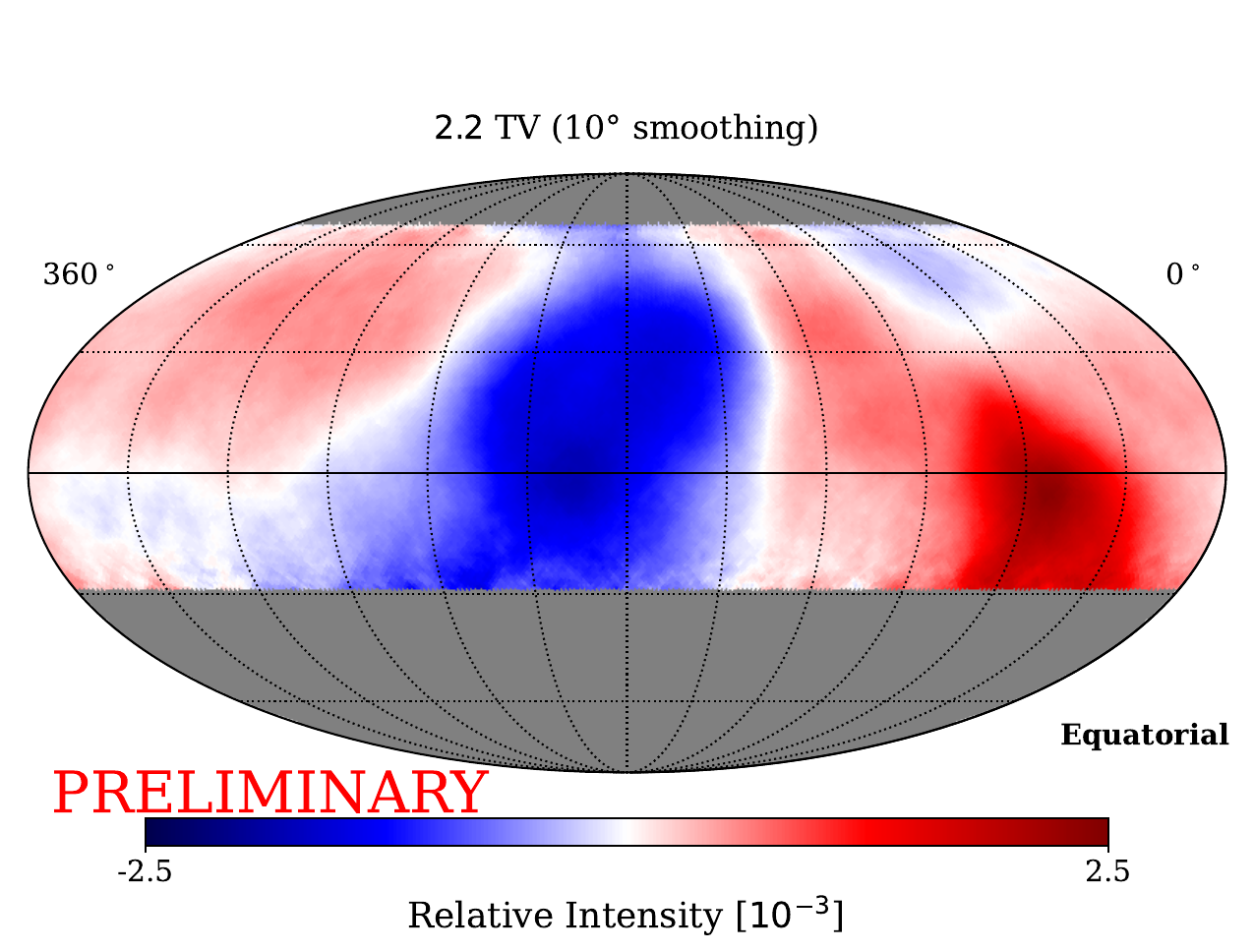}
  \includegraphics[width=0.30\textwidth]{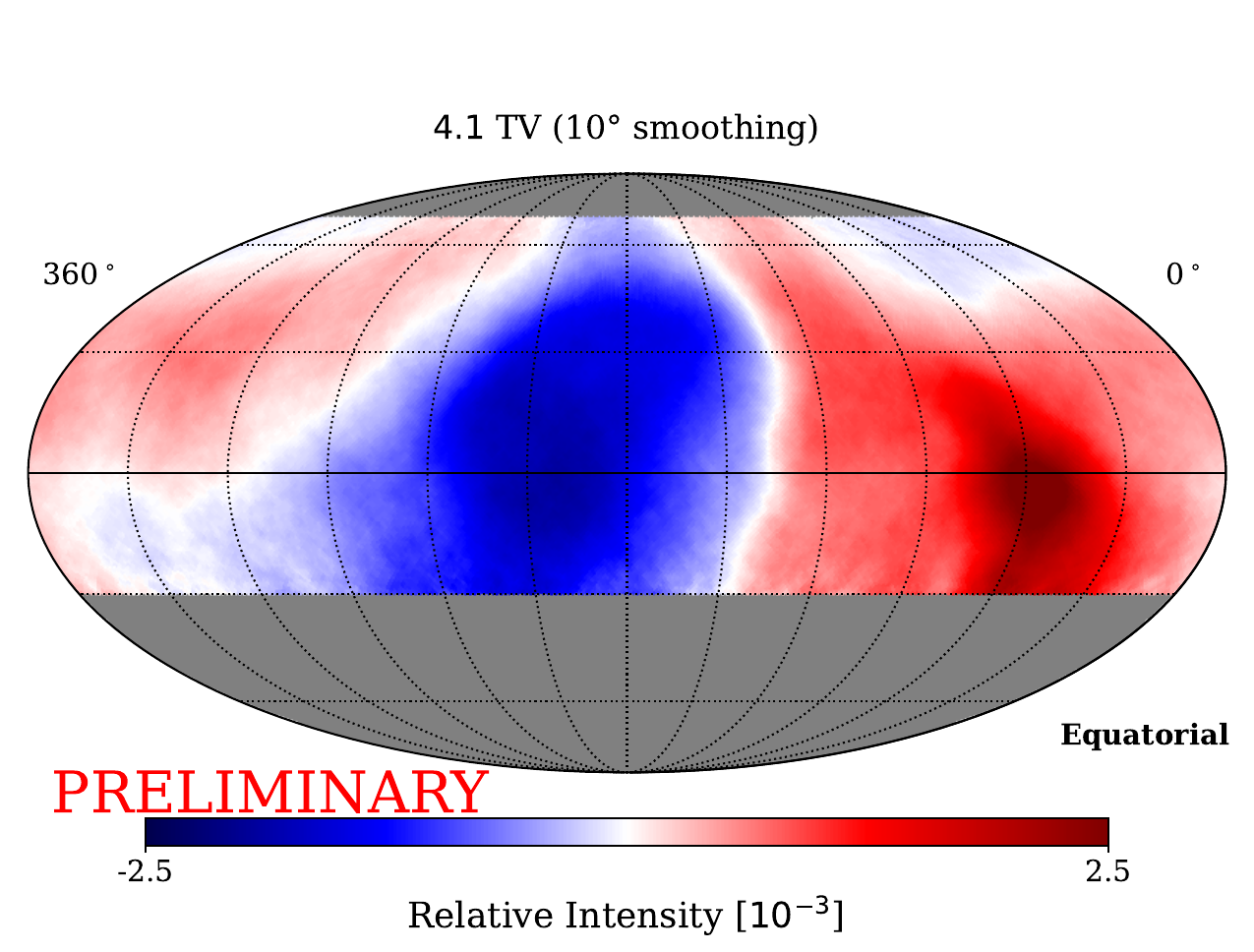}
  \includegraphics[width=0.30\textwidth]{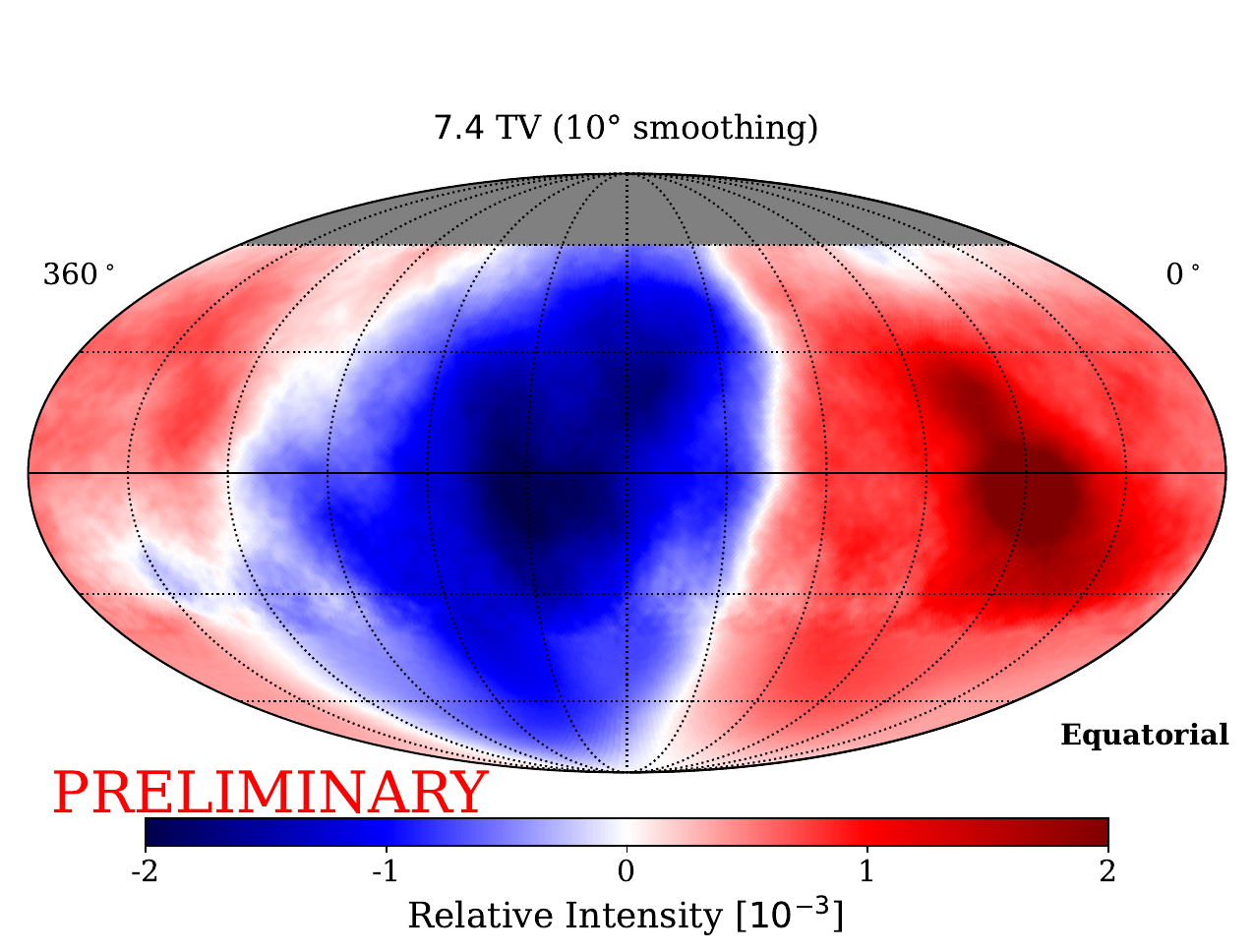}
  \includegraphics[width=0.30\textwidth]{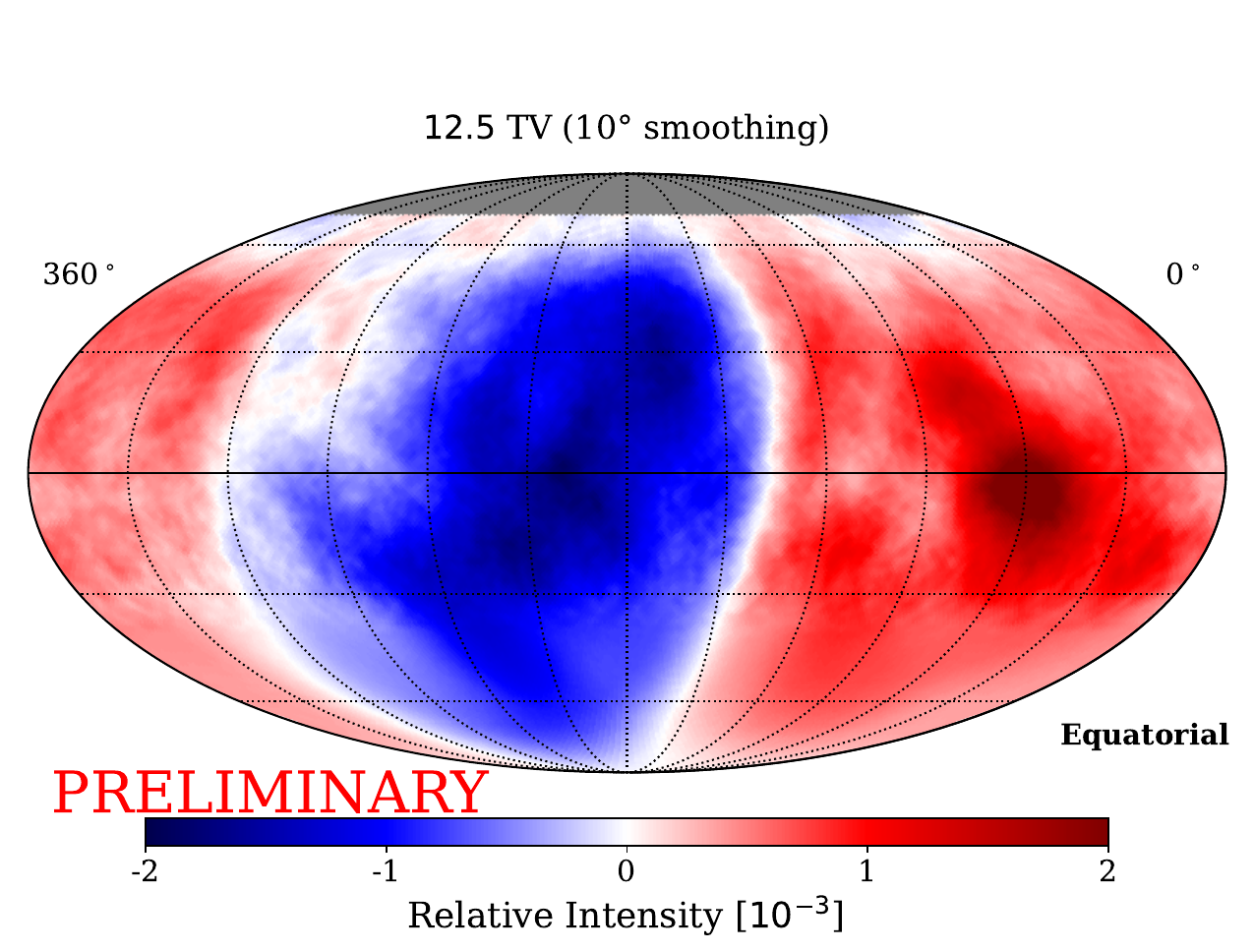}
  \includegraphics[width=0.30\textwidth]{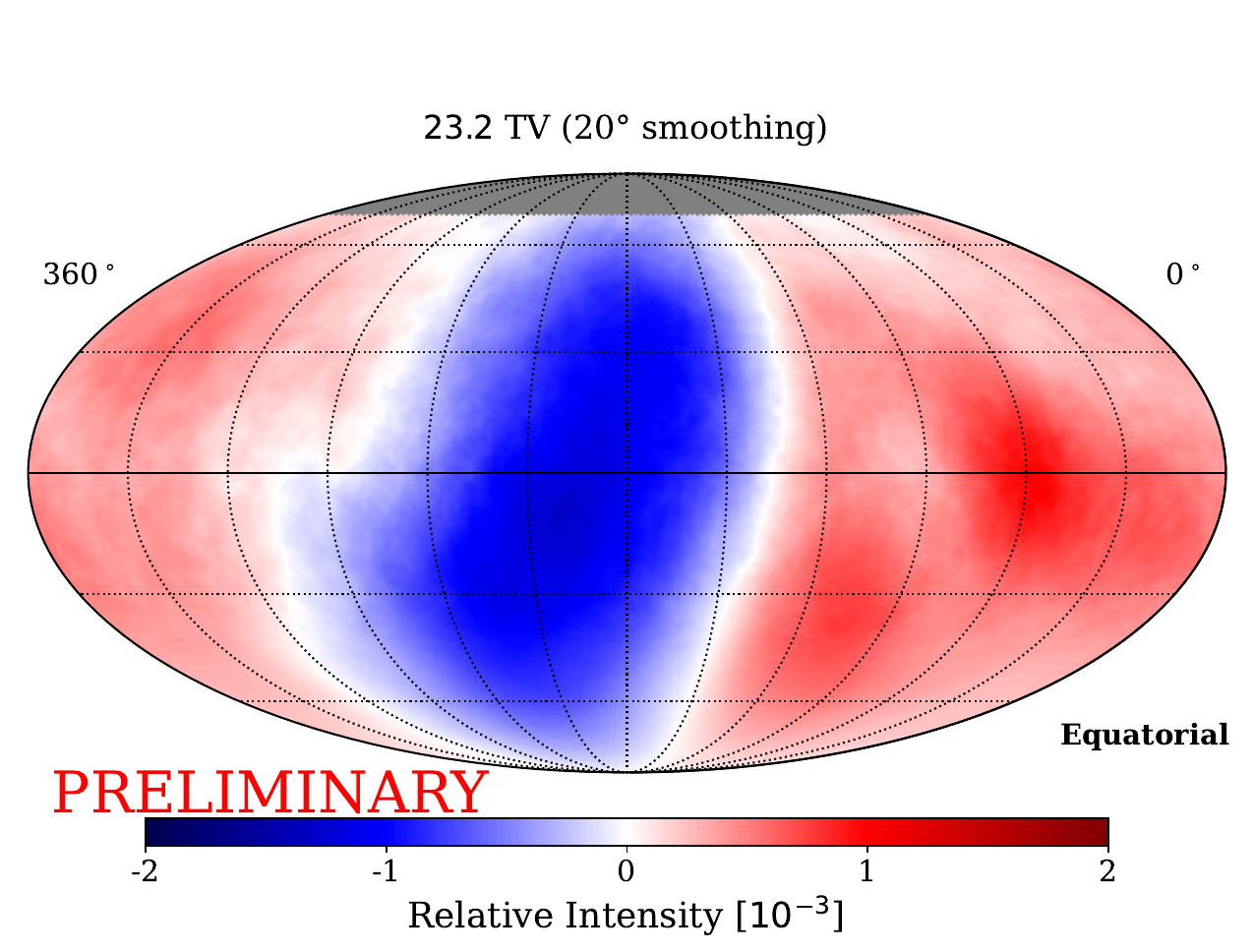}
  \includegraphics[width=0.30\textwidth]{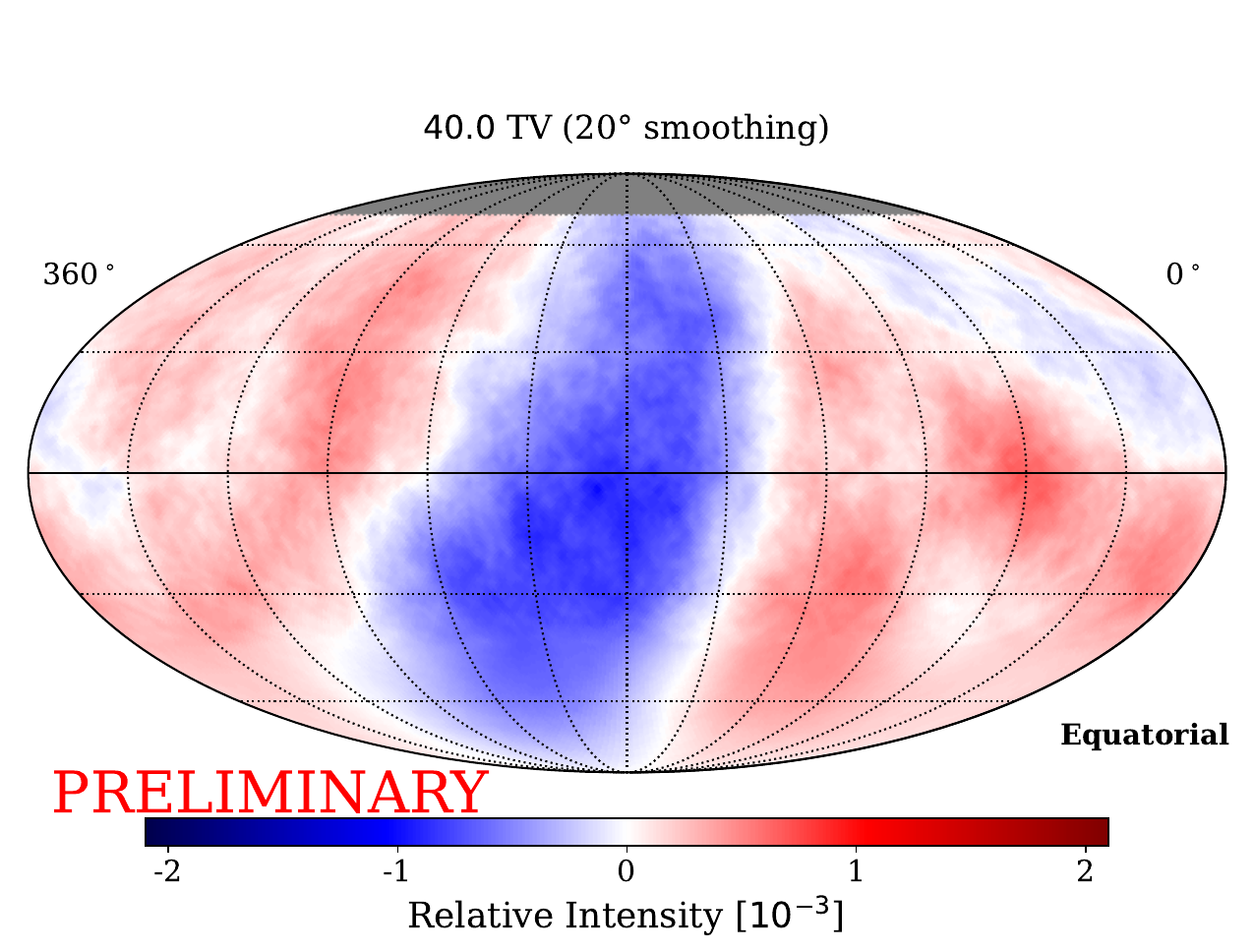}
  \includegraphics[width=0.30\textwidth]{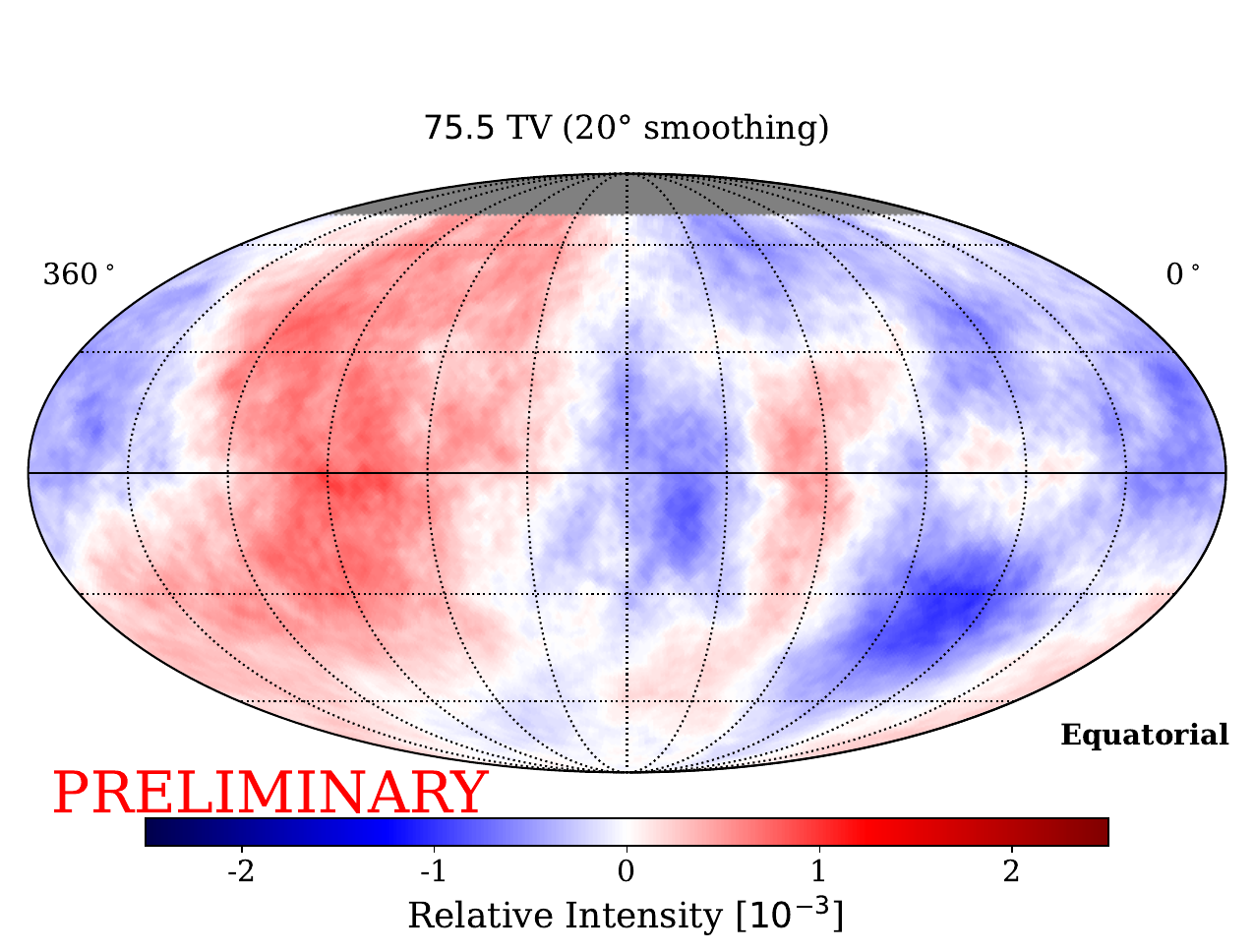}
  \includegraphics[width=0.30\textwidth]{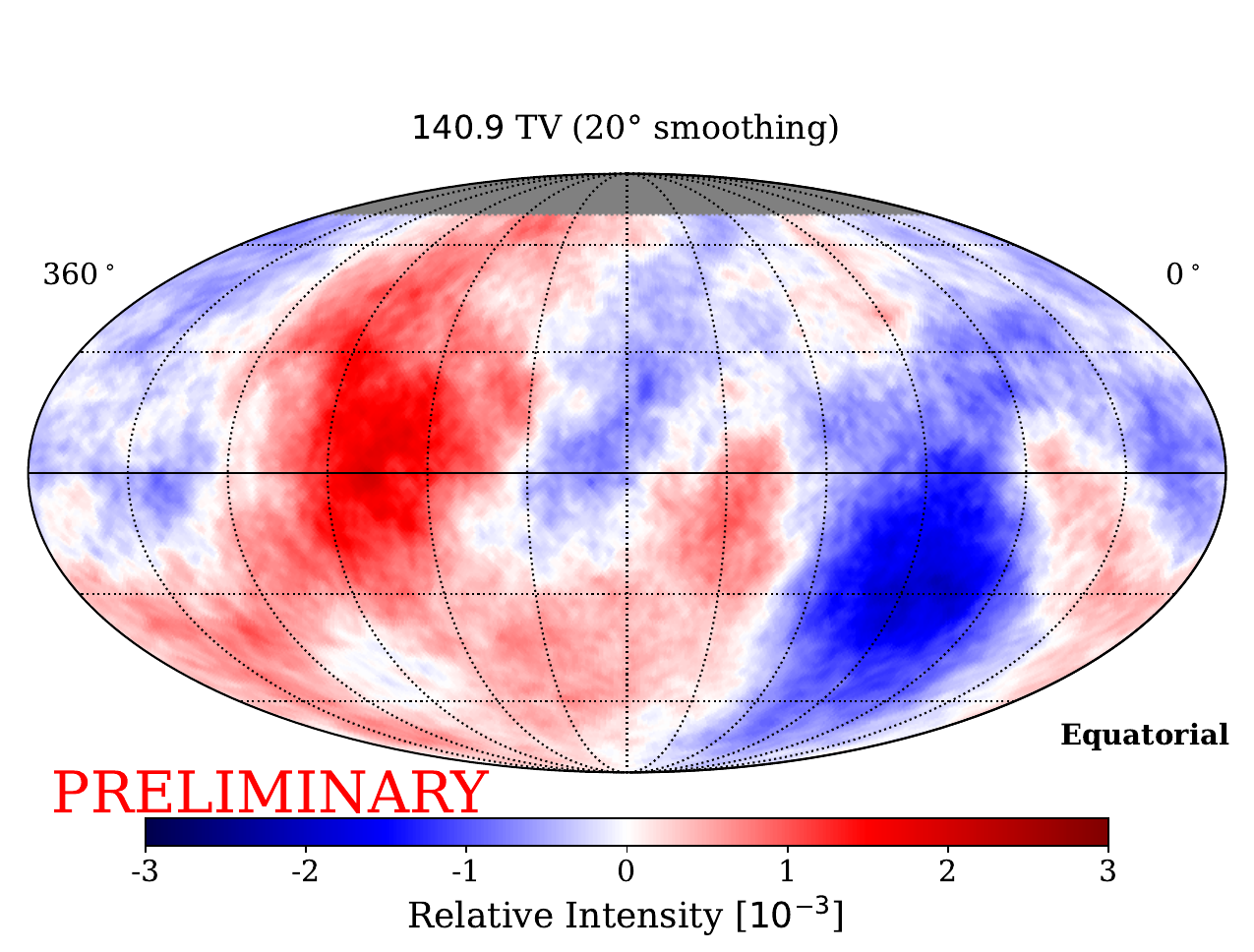}
  \includegraphics[width=0.30\textwidth]{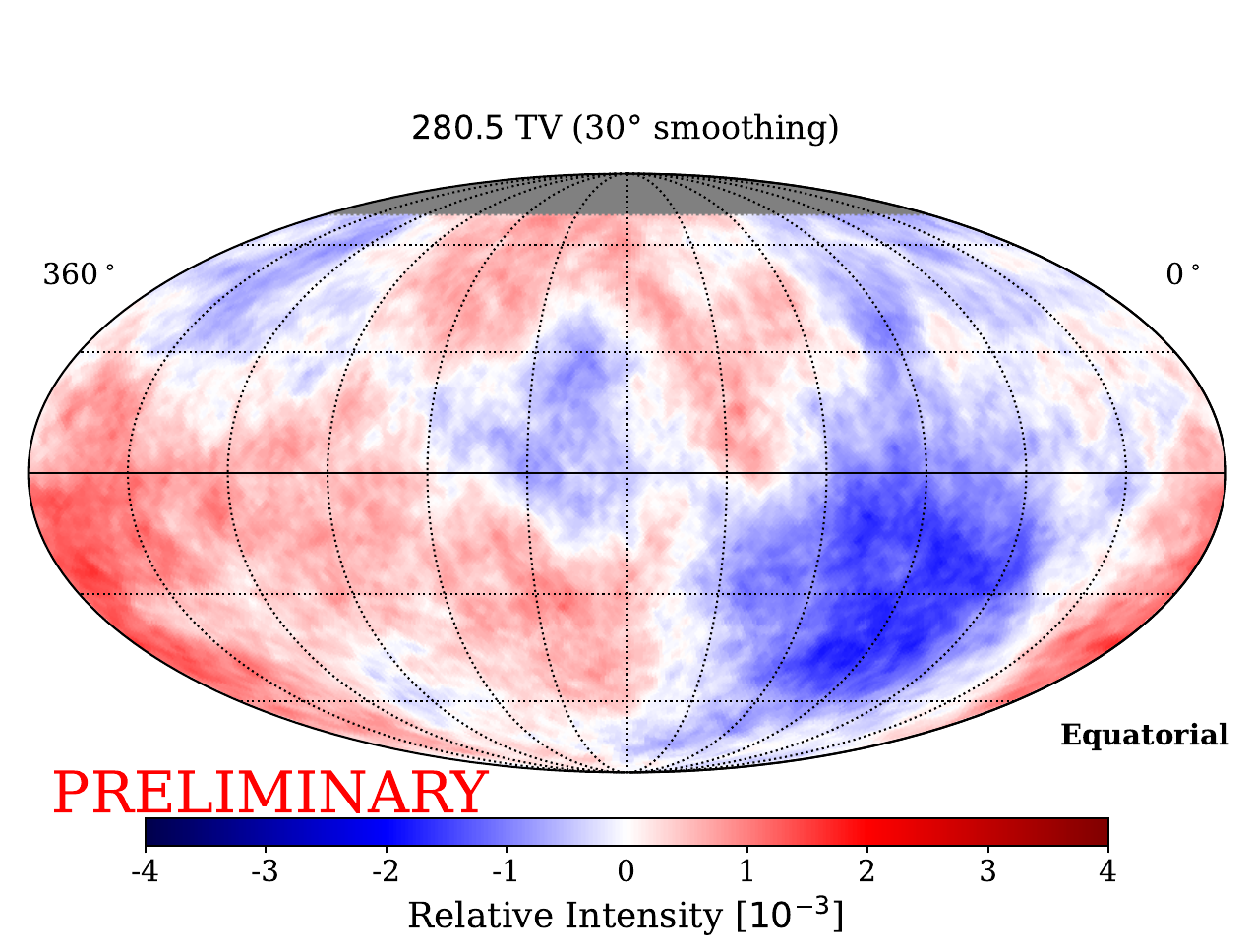}
  \caption{\footnotesize Relative intensity for 11 HAWC energy bins. Table \ref{tab:energy_bins} shows the median energies corresponding to each bin. The last seven maps correspond to rigidity-matching pairs of energy bins. Combined relative intensity maps are reconstructed using the maximum likelihood method of Ahlers et al.~\cite{0004-637X-823-1-10}. A rapid phase transition is observed going from 40 TV to 76 TV, consistent with previous individual measurements.}
  \label{fig:combined_ri}
\end{figure*}

\begin{figure*}[hbtp]
  \centering
  \includegraphics[width=0.30\textwidth]{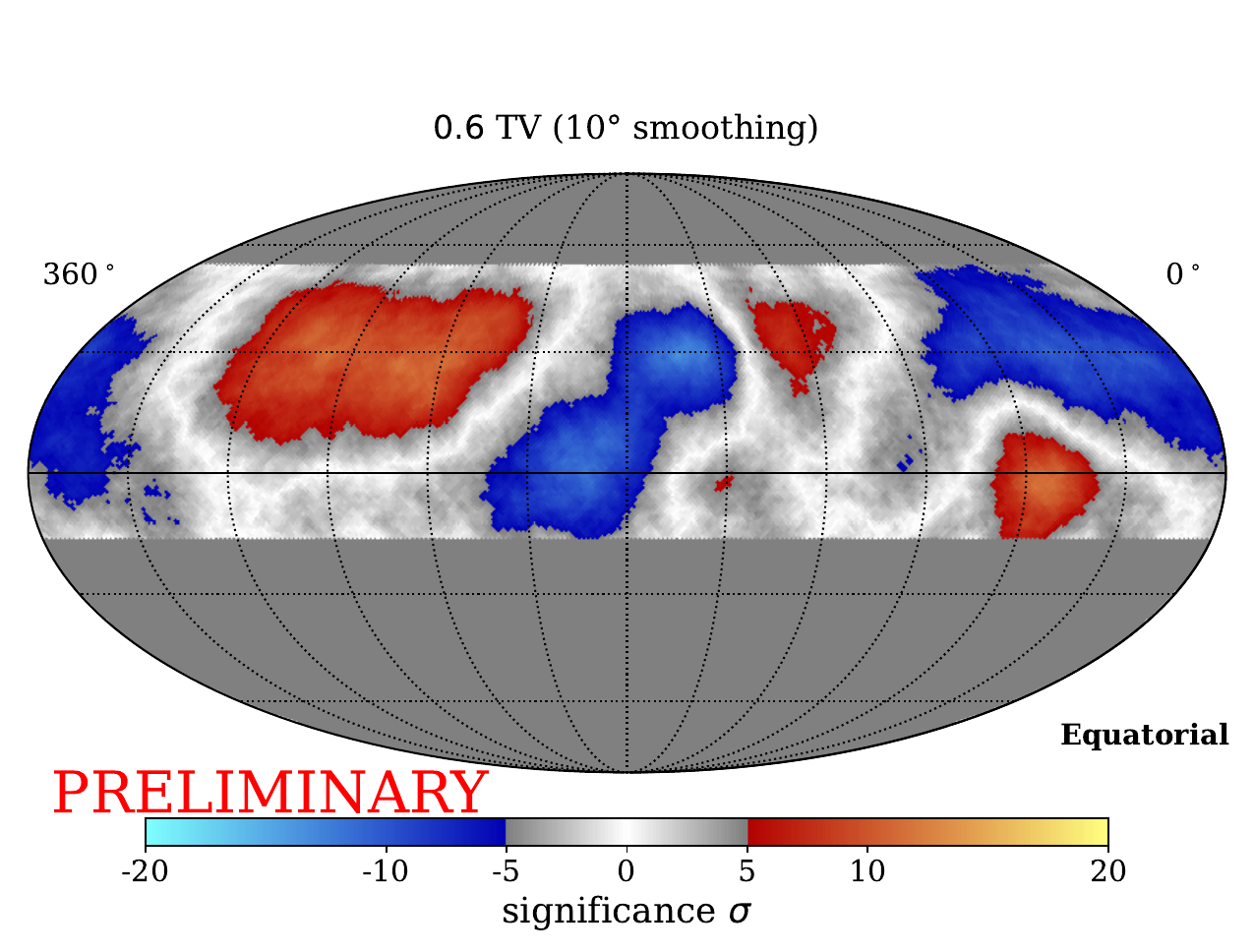}
  \includegraphics[width=0.30\textwidth]{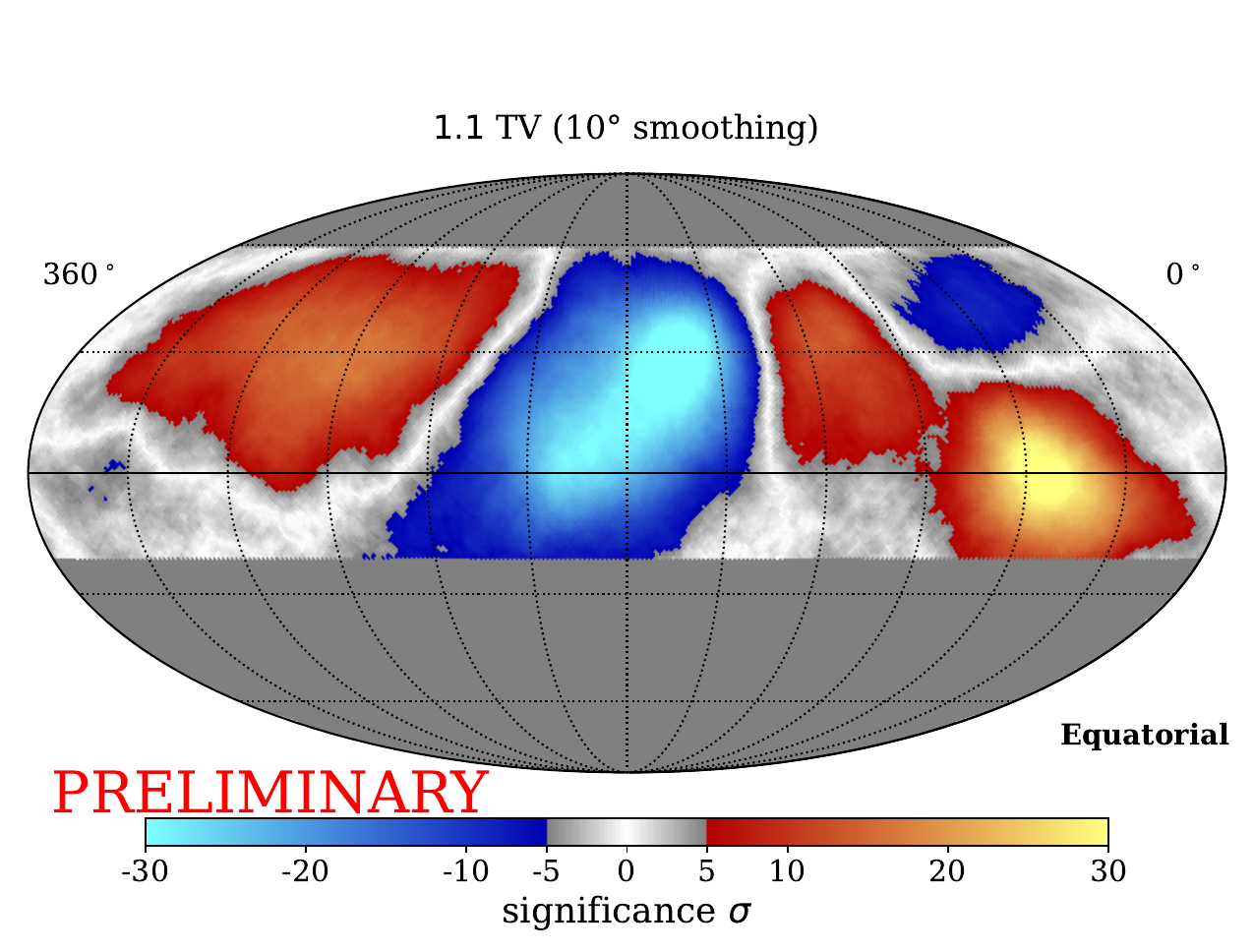}
  \includegraphics[width=0.30\textwidth]{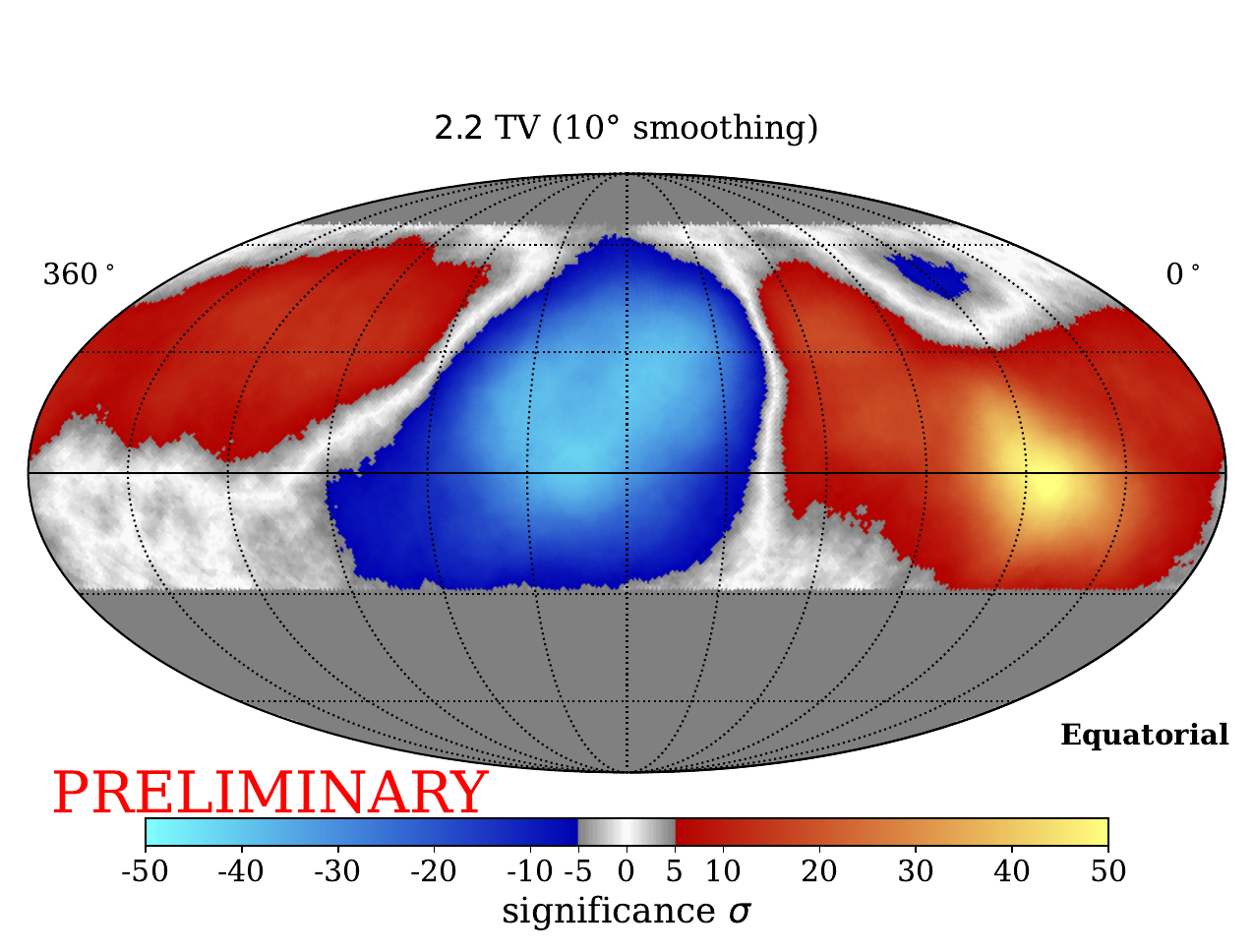}
  \includegraphics[width=0.30\textwidth]{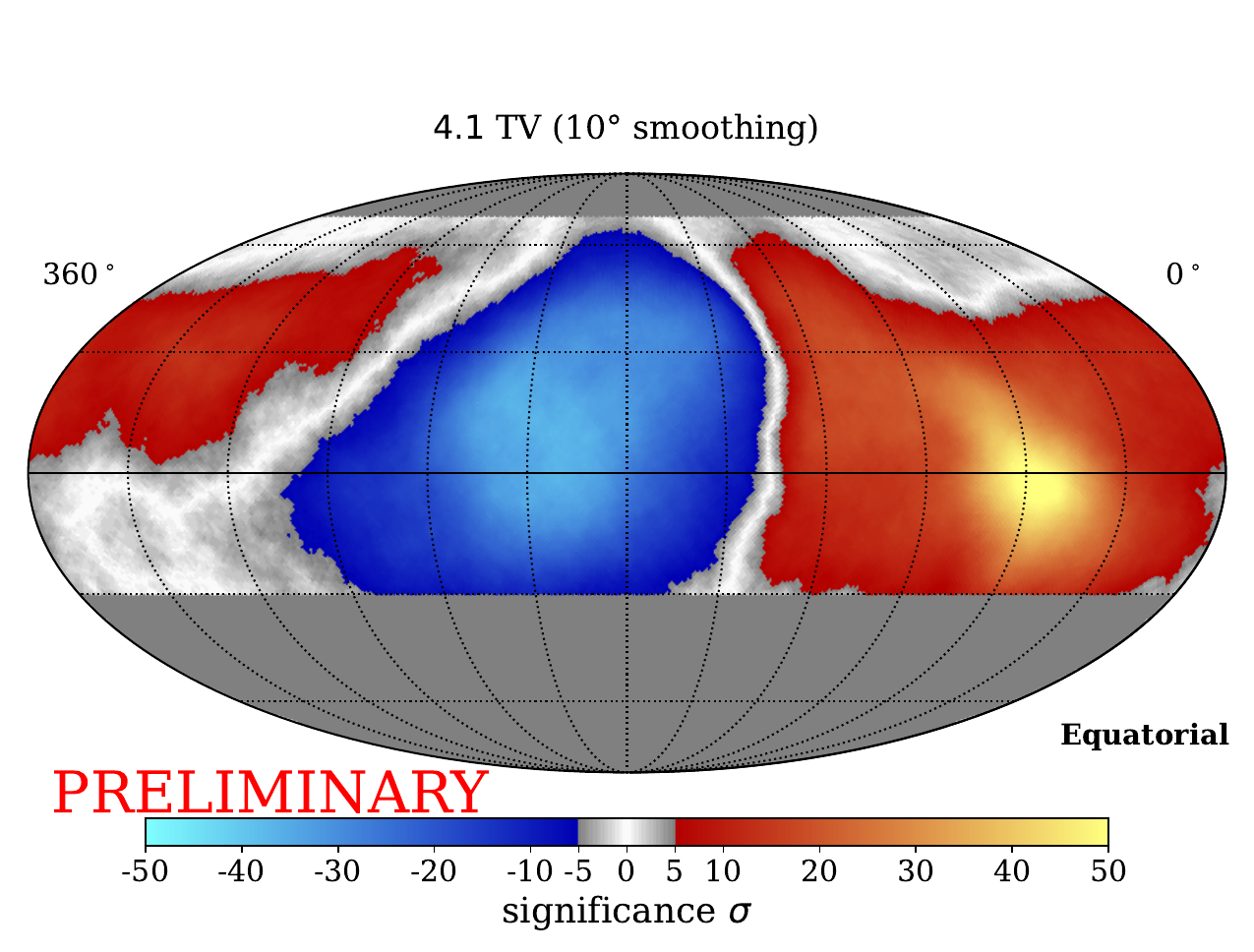}
  \includegraphics[width=0.30\textwidth]{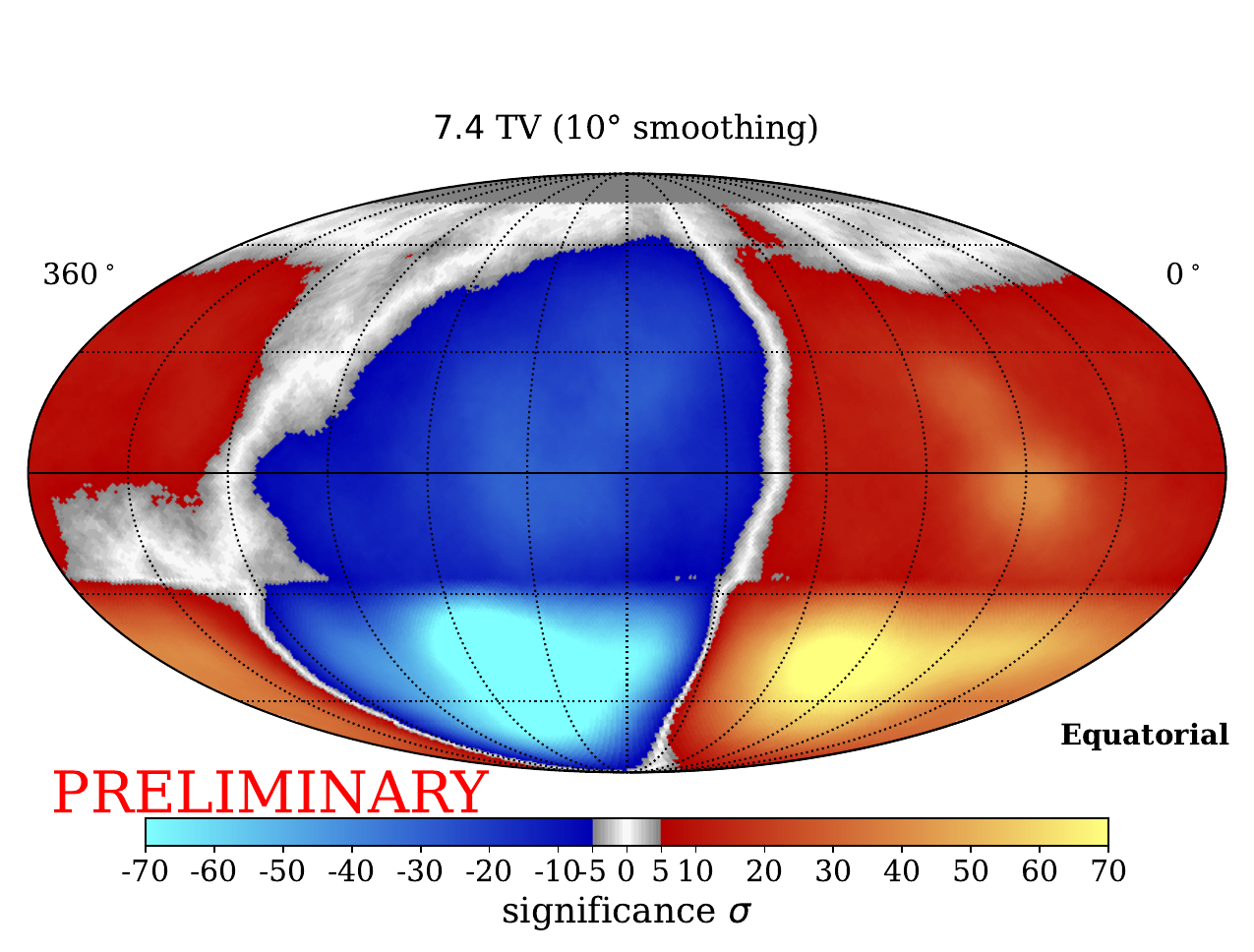}
  \includegraphics[width=0.30\textwidth]{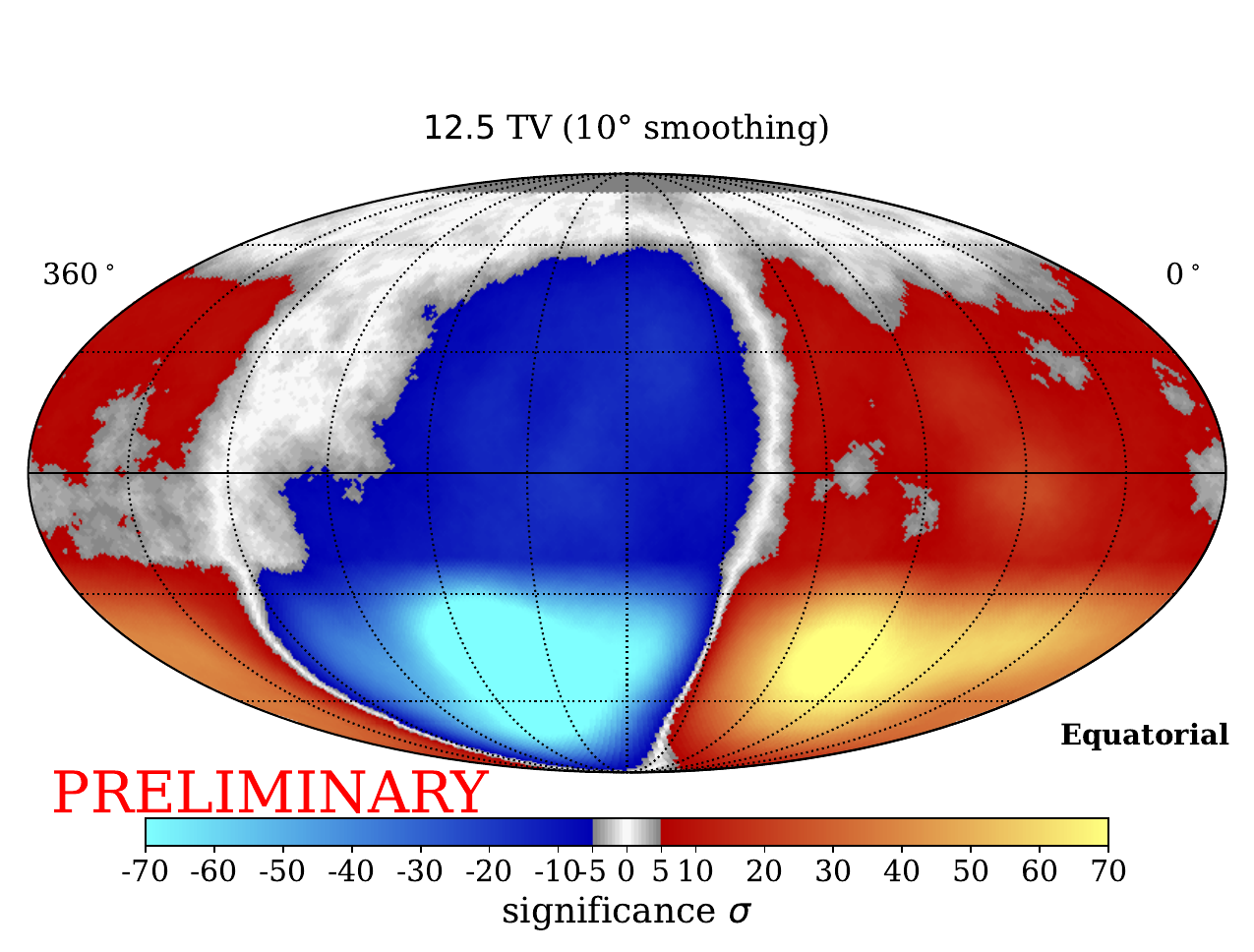}
  \includegraphics[width=0.30\textwidth]{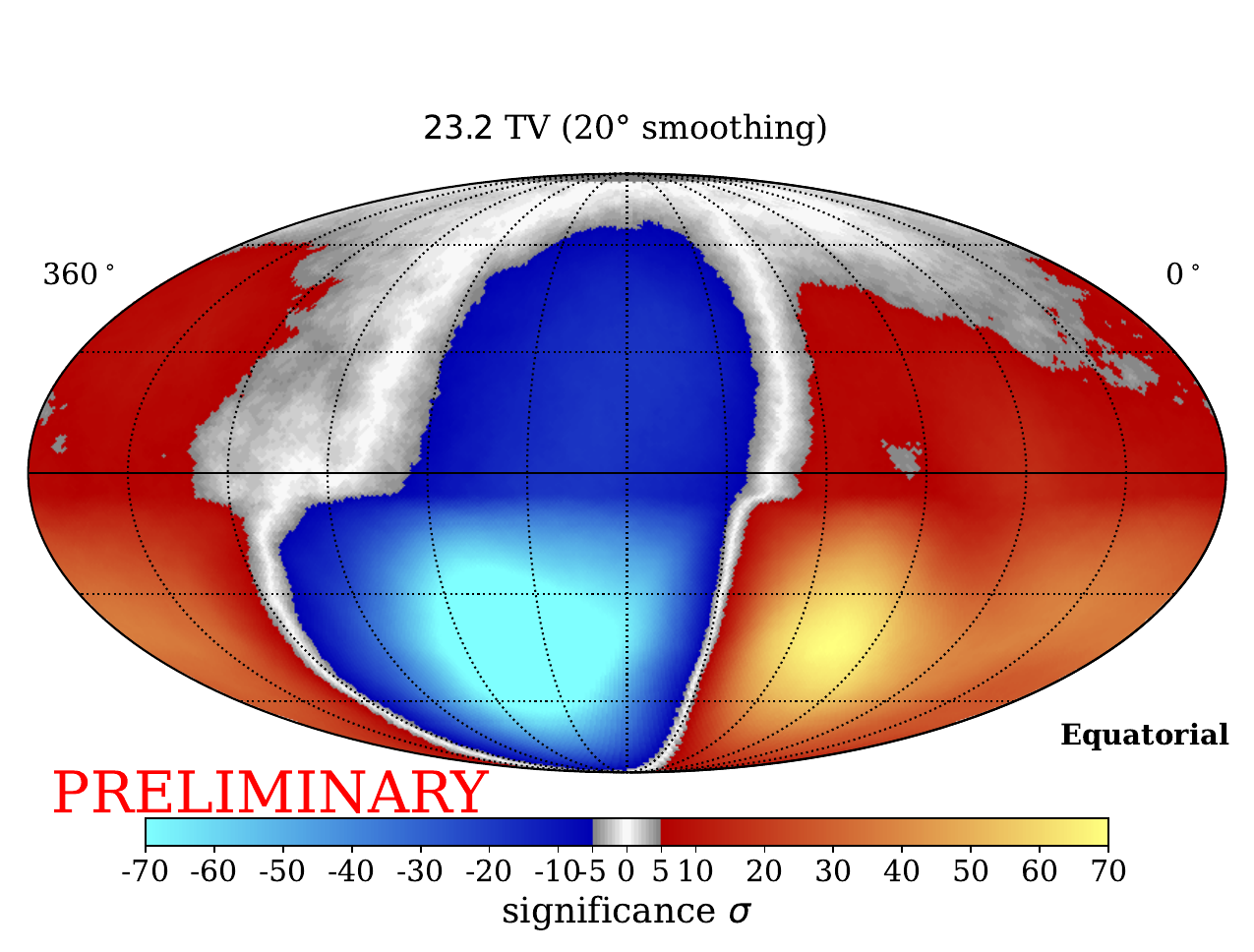}
  \includegraphics[width=0.30\textwidth]{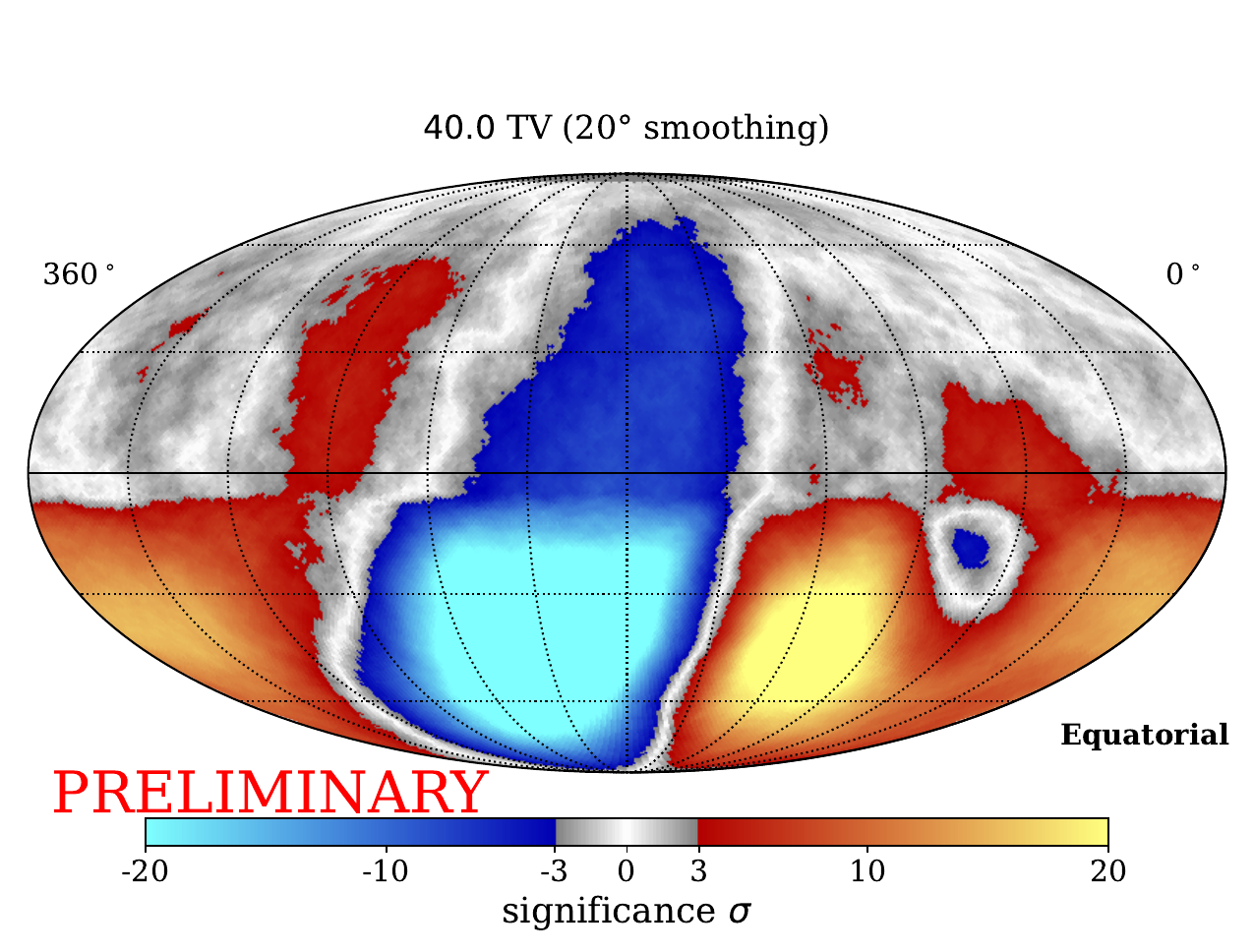}
  \includegraphics[width=0.30\textwidth]{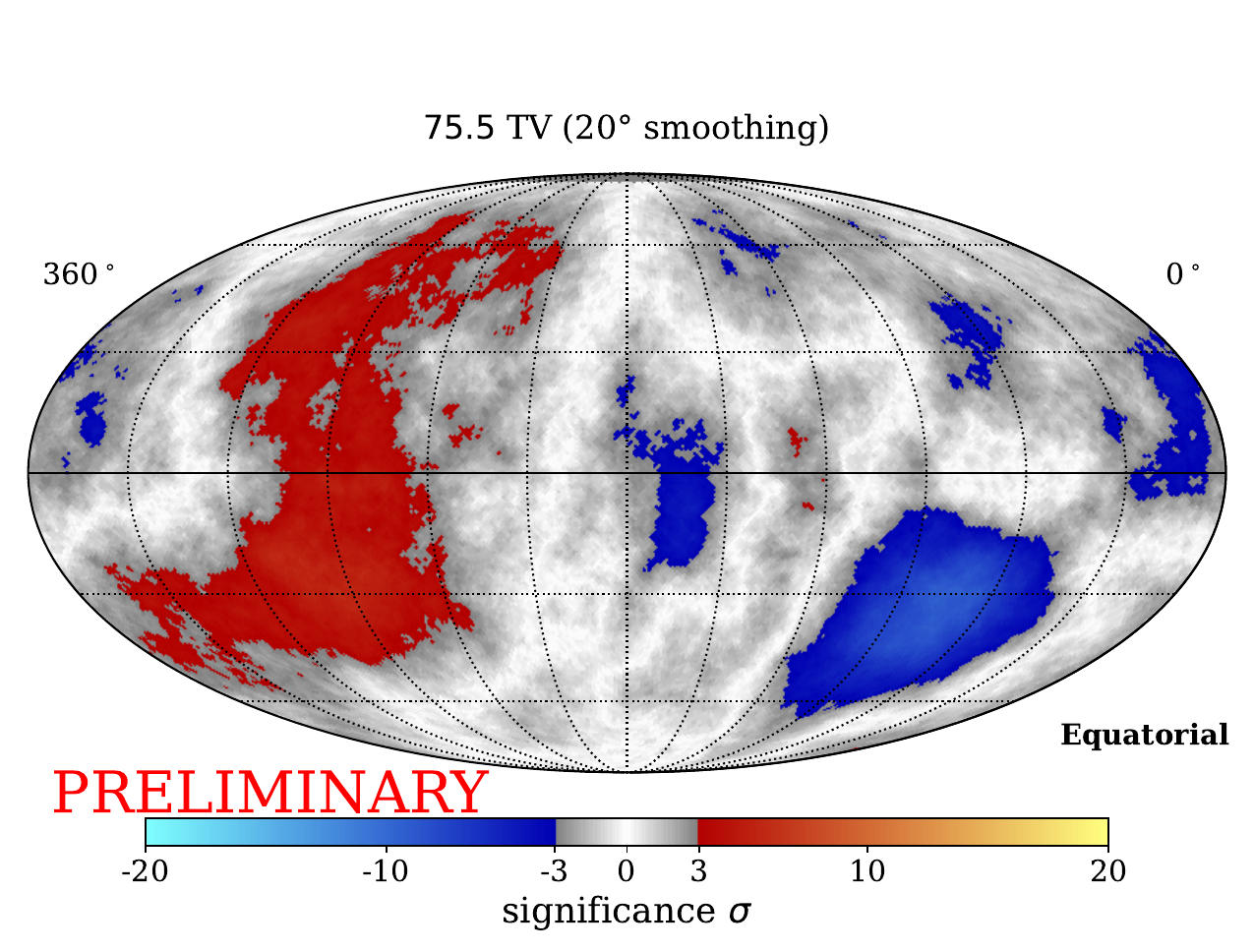}
  \includegraphics[width=0.30\textwidth]{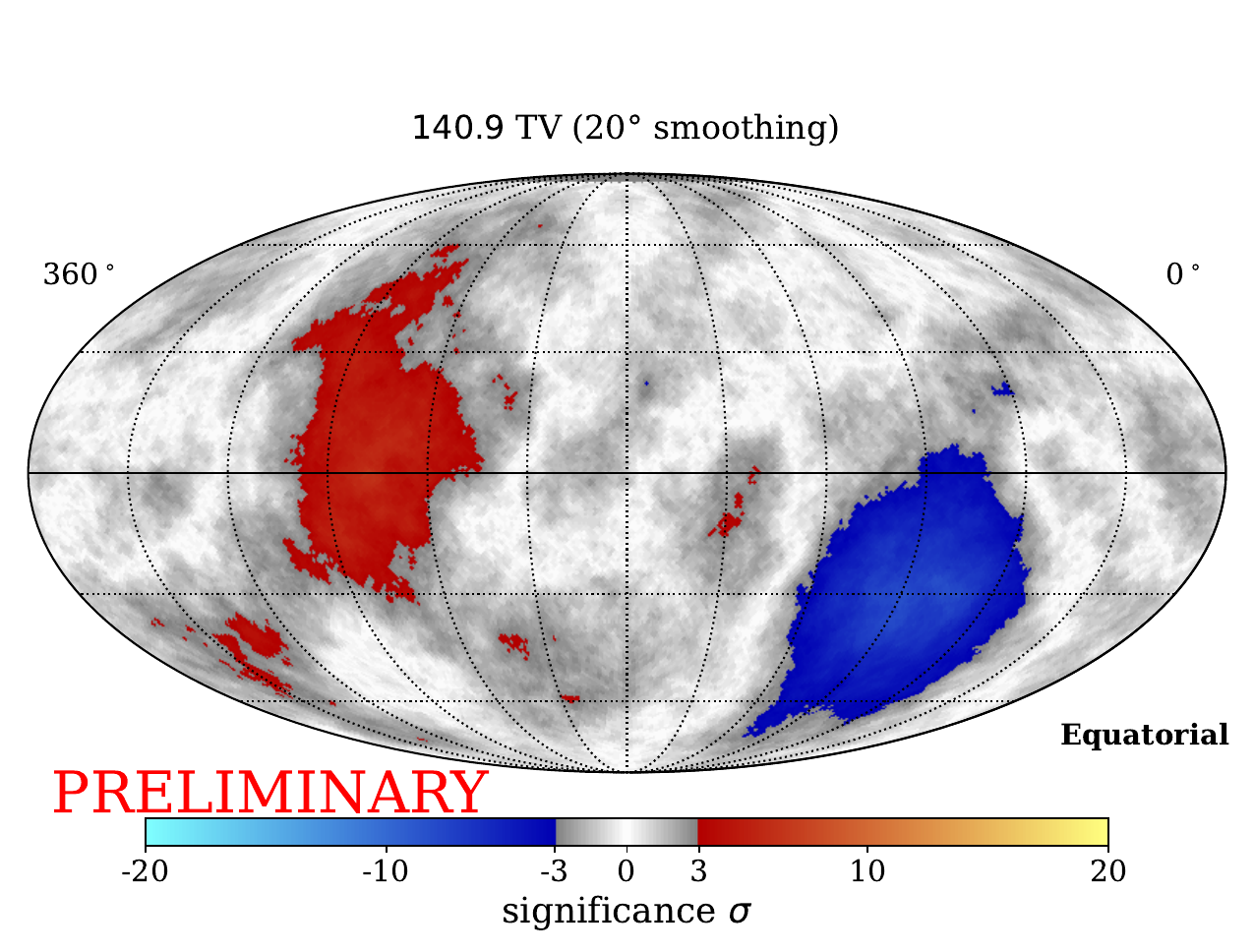}
  \includegraphics[width=0.30\textwidth]{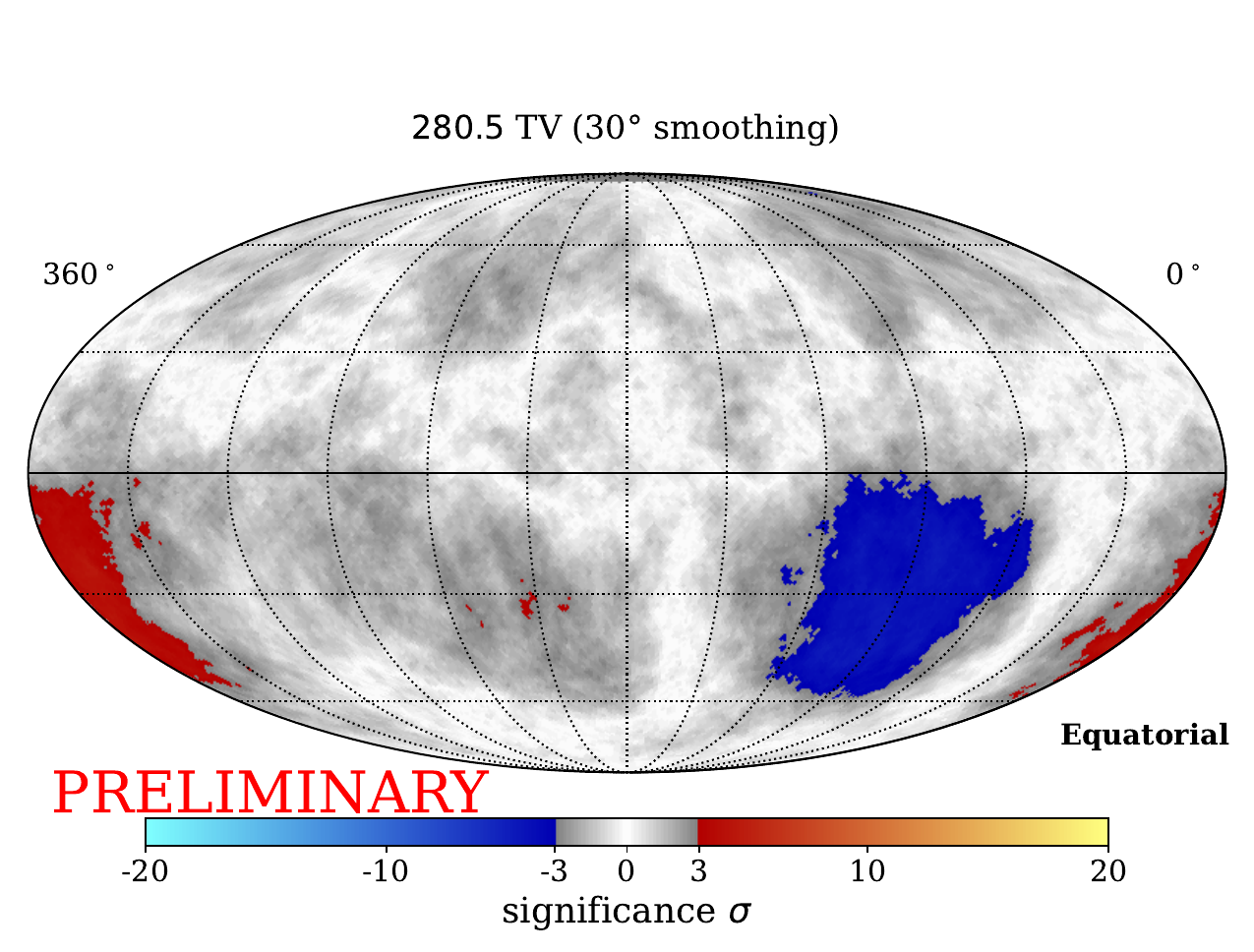}
  \caption{\footnotesize Li--Ma significance for 11 HAWC energy bins. Table \ref{tab:energy_bins} shows the median energies corresponding to each bin. The last seven maps correspond to rigidity-matching pairs of energy bins. Smoothing radius and thresholds are adjusted for higher energy bins to compensate for decreasing statistics.}
  \label{fig:combinedsig}
\end{figure*}

If we assume that the distribution of arrival directions of cosmic rays is rigidity-dependent rather than energy-dependent, we can compare the one-dimensional distributions for IceCube and HAWC at comparable rigidities and using a $\chi^2$ test and assess whether we achieve better agreement. 
In order to find the optimal IceCube energy cuts to match the rigidity for each HAWC bin, we perform a scan and use a Kolmogorov-Smirnov test to compare the rigidity distributions assuming the GSF~\citep{Dembinski:2017N7} composition model, as shown in Fig~\ref{fig:mc_logz}. Table \ref{tab:energy_bins} shows the median energies and rigidity for each bin.
The combined IceCube--HAWC maps are shown in relative intensity (Fig.~\ref{fig:combined_ri}) and significance (Fig~\ref{fig:combinedsig}) for seven rigidity-driven pairs of energy bins. The smoothing radius and thresholds are adjusted for higher energy bins to compensate for the decreasing statistics. A rapid phase transition is observed going from 40~TV to 76~TV consistent with previous individual measurements~\citep{Abbasi_2025}.

\subsection{Angular Power Spectrum}\label{sec:power_spectrum}
\begin{figure*}[hbtp]
  \centering
  \includegraphics[width=0.31\textwidth]{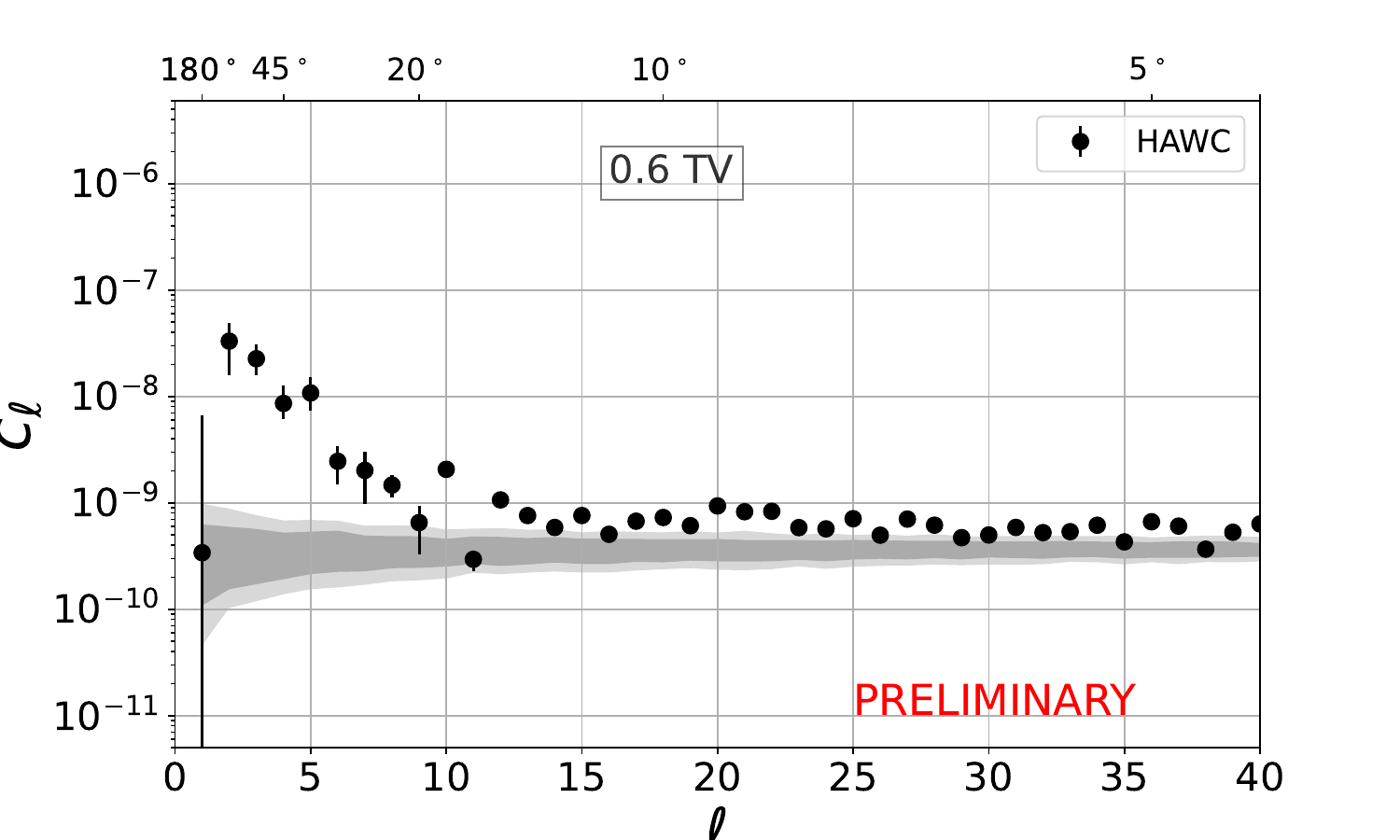}
  \includegraphics[width=0.31\textwidth]{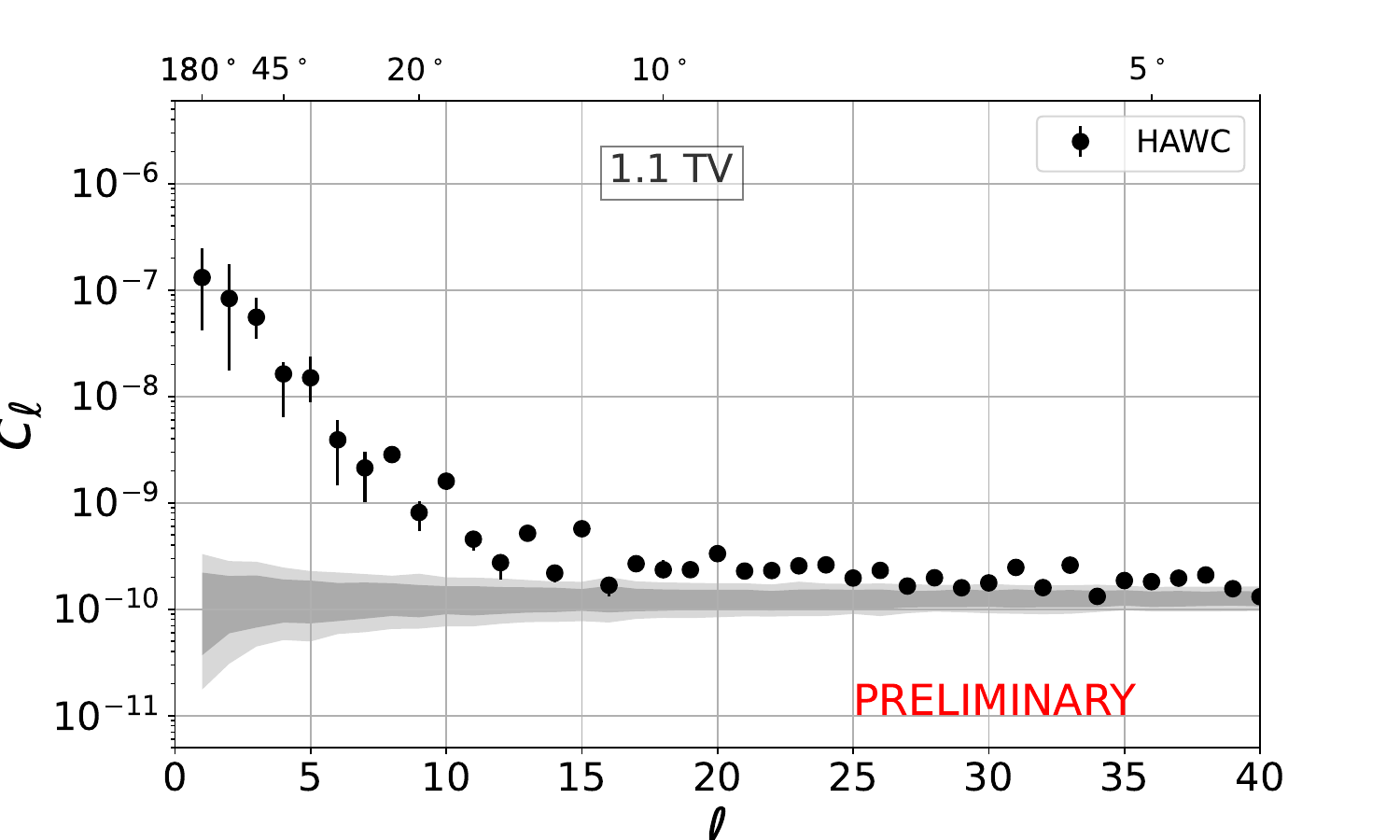}
  \includegraphics[width=0.31\textwidth]{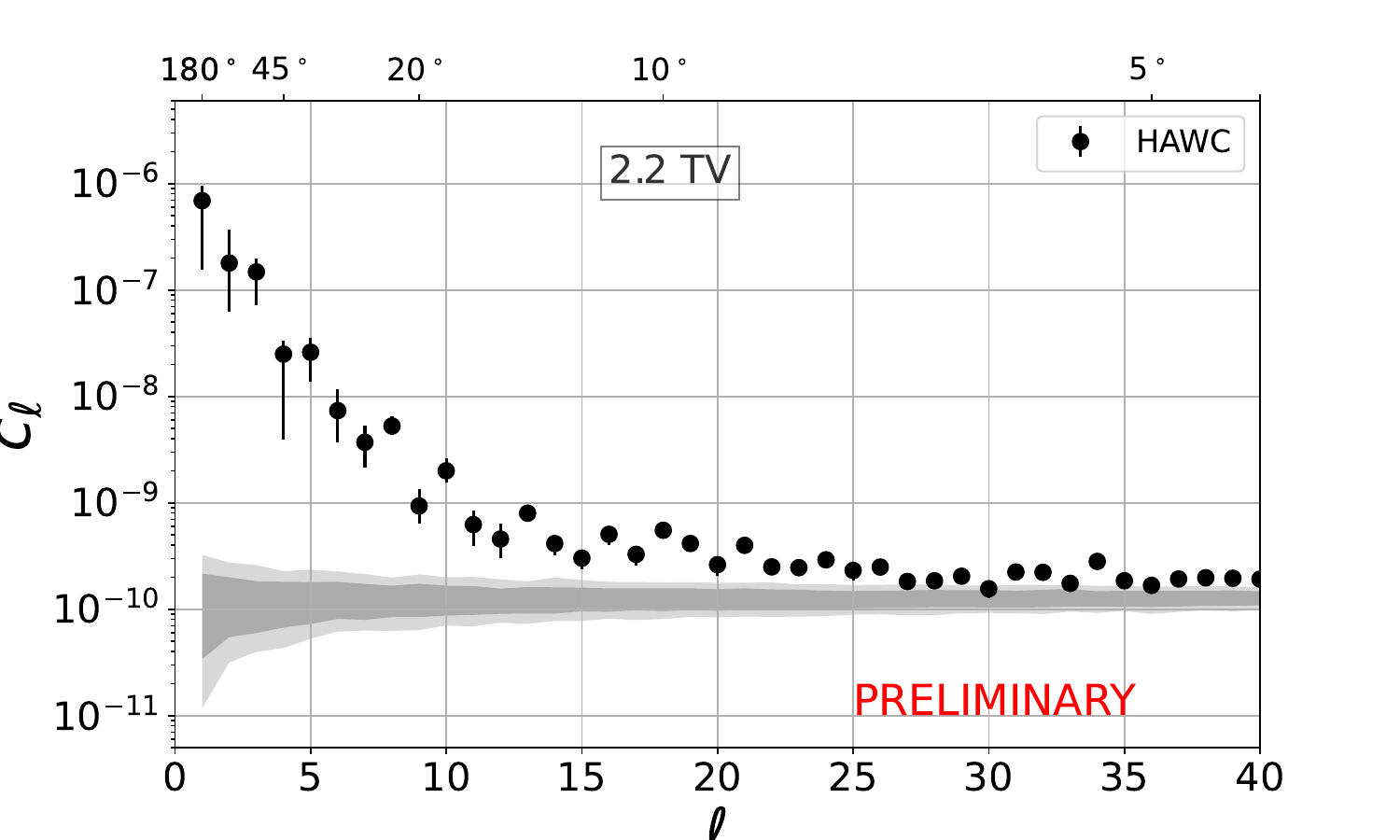}
  \includegraphics[width=0.31\textwidth]{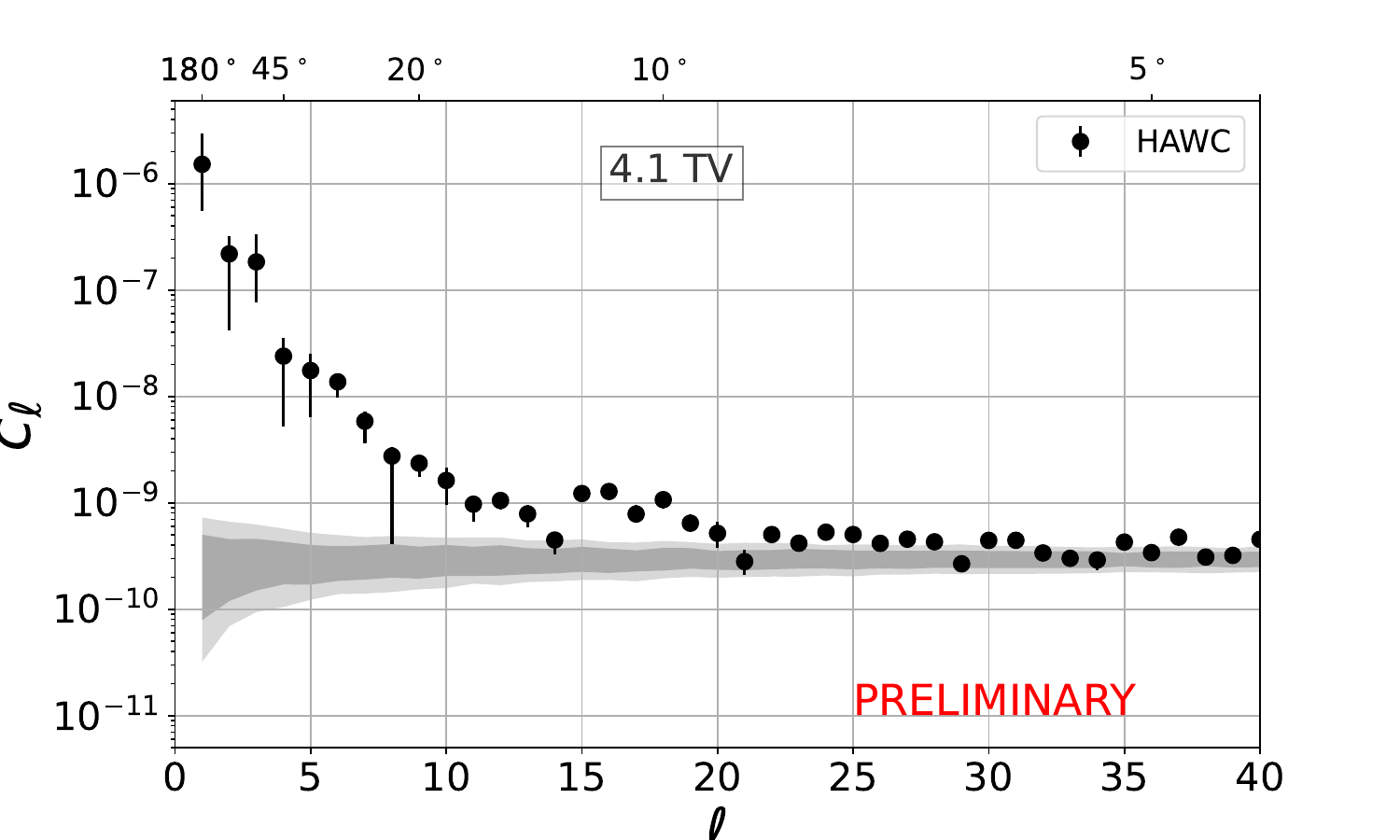}
  \includegraphics[width=0.31\textwidth]{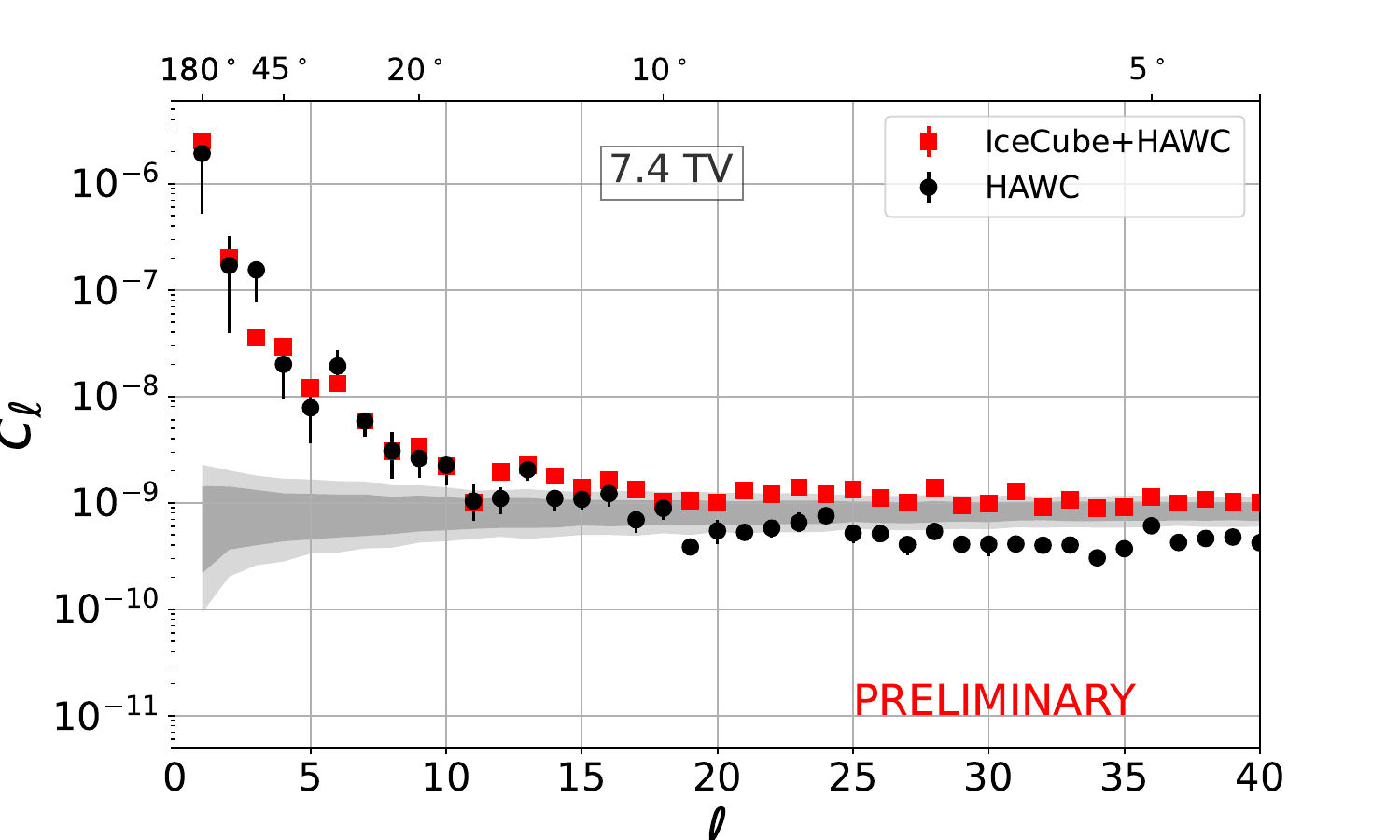}
  \includegraphics[width=0.31\textwidth]{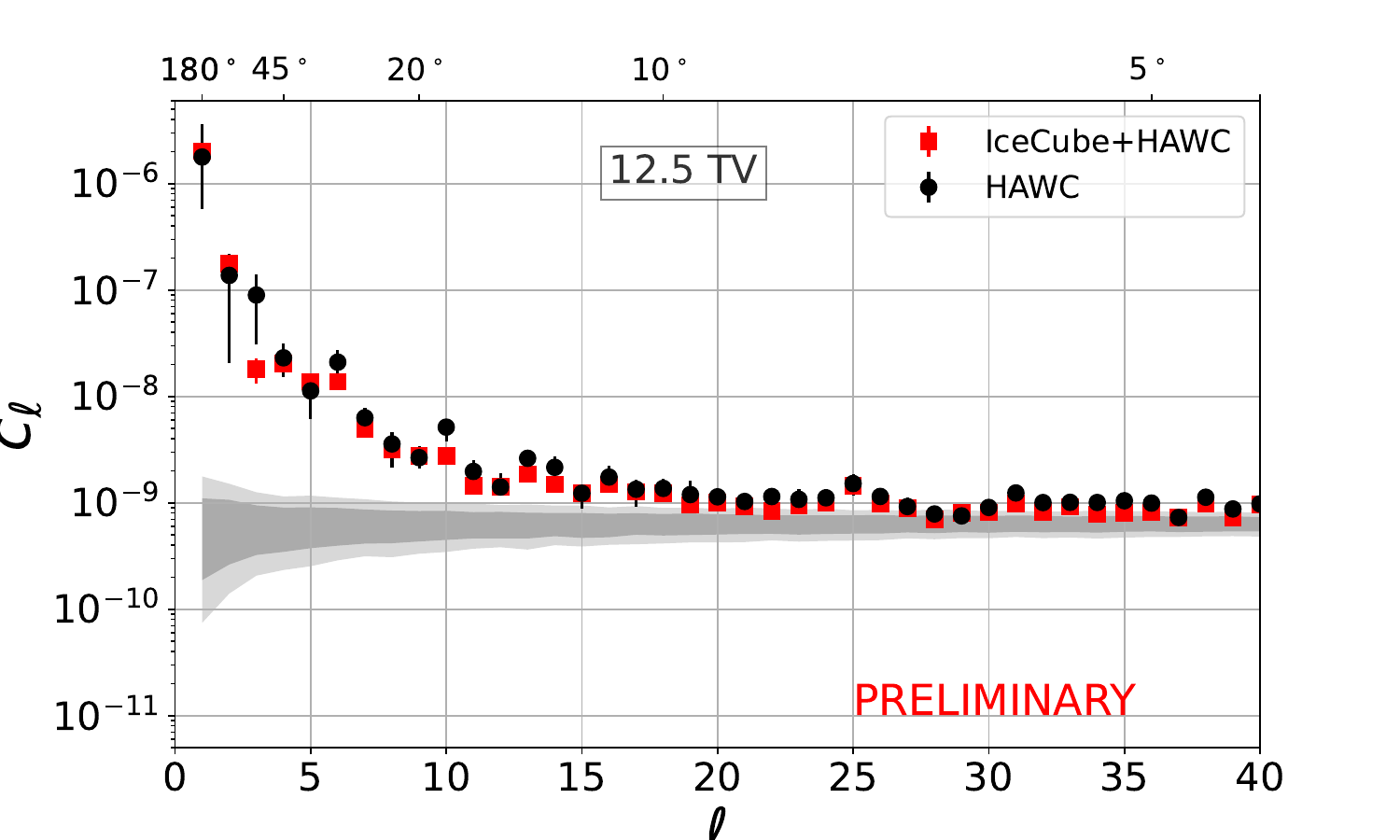}
  \includegraphics[width=0.31\textwidth]{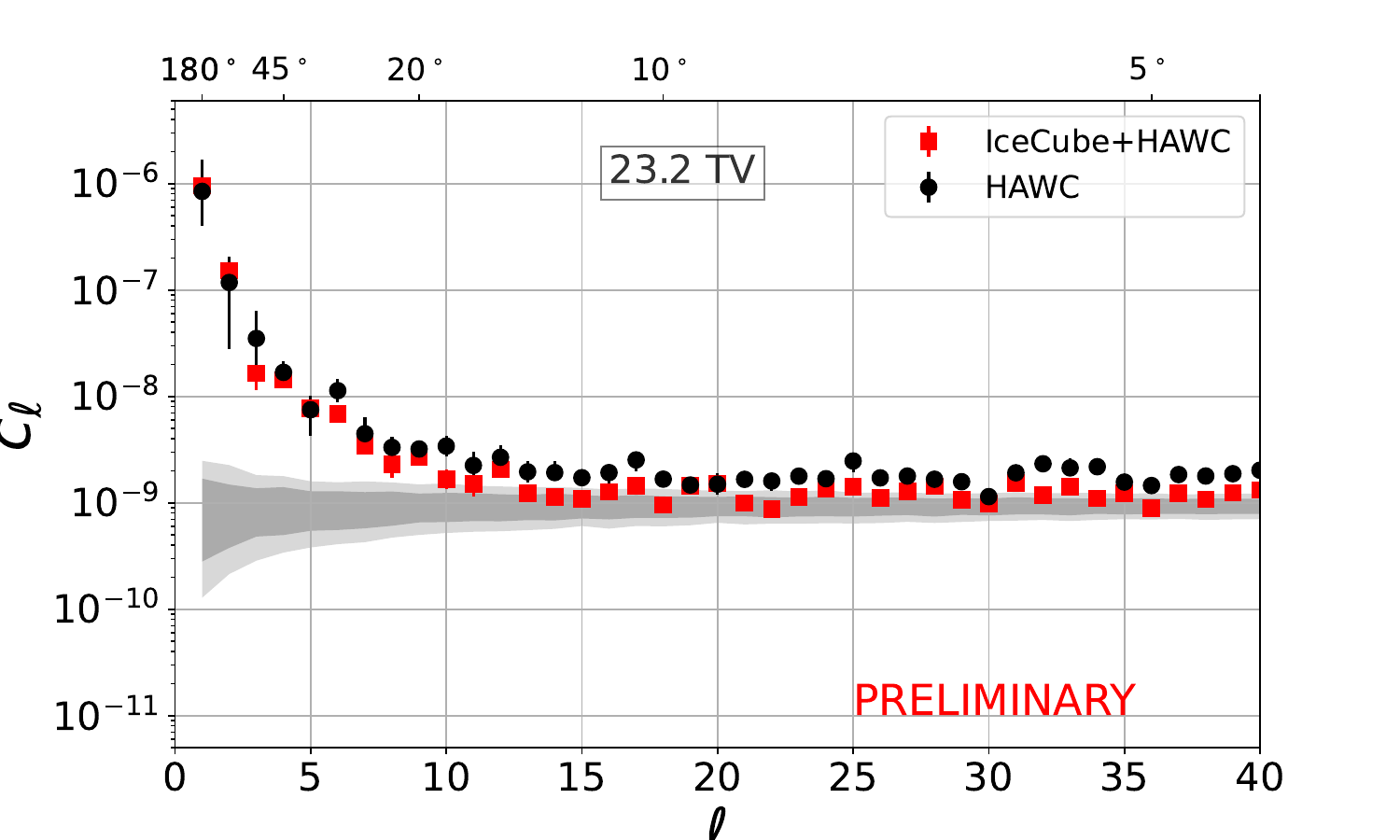}
  \includegraphics[width=0.31\textwidth]{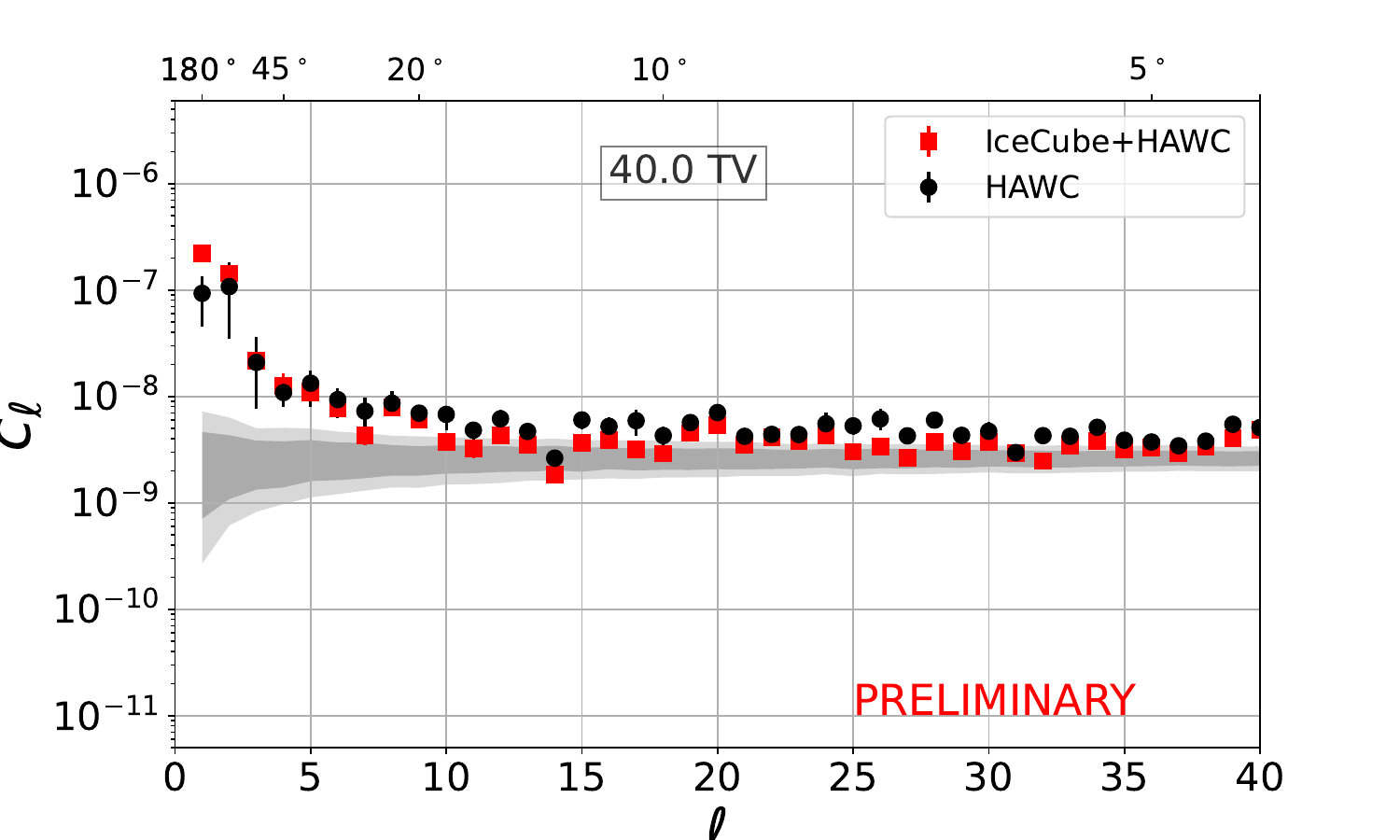}
  \includegraphics[width=0.31\textwidth]{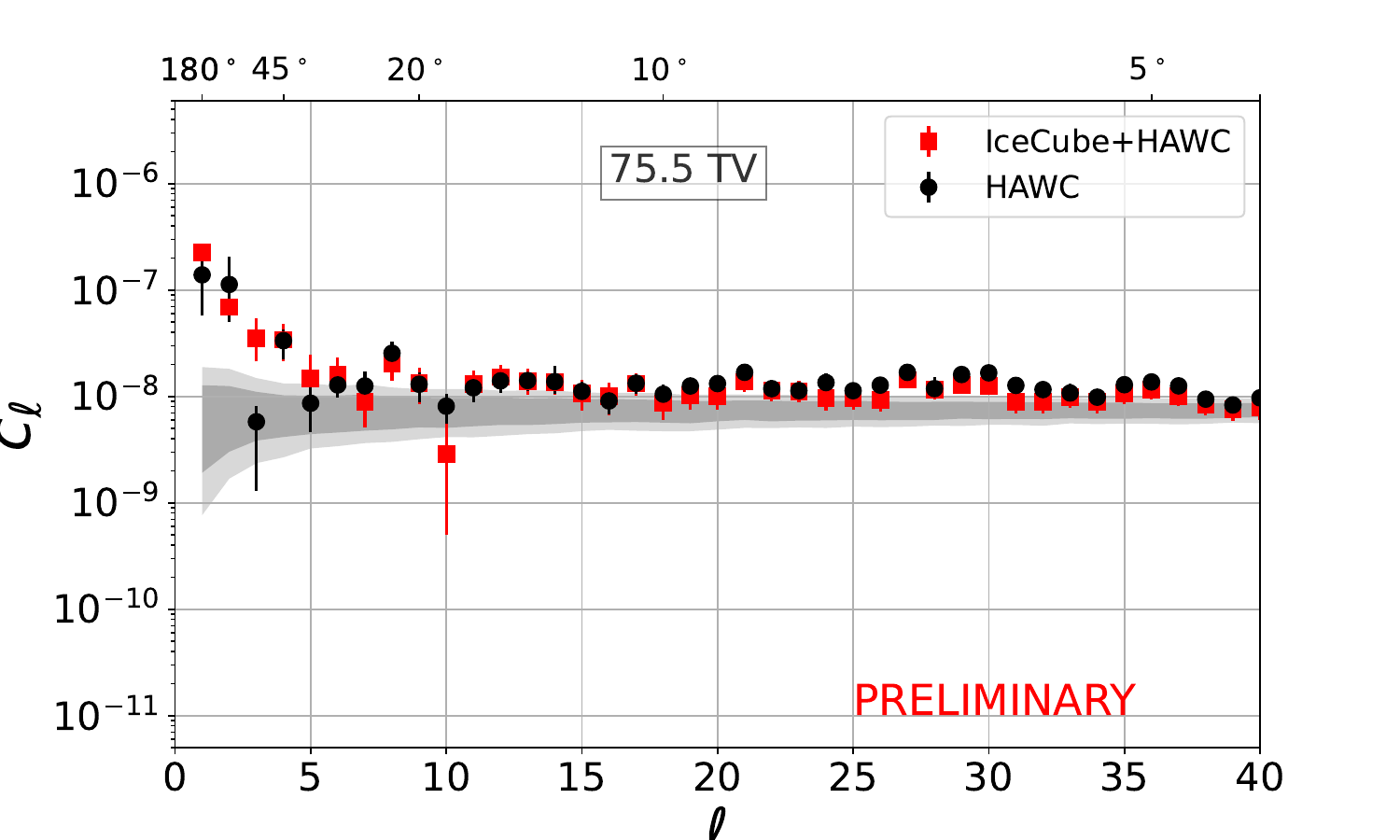}
  \includegraphics[width=0.31\textwidth]{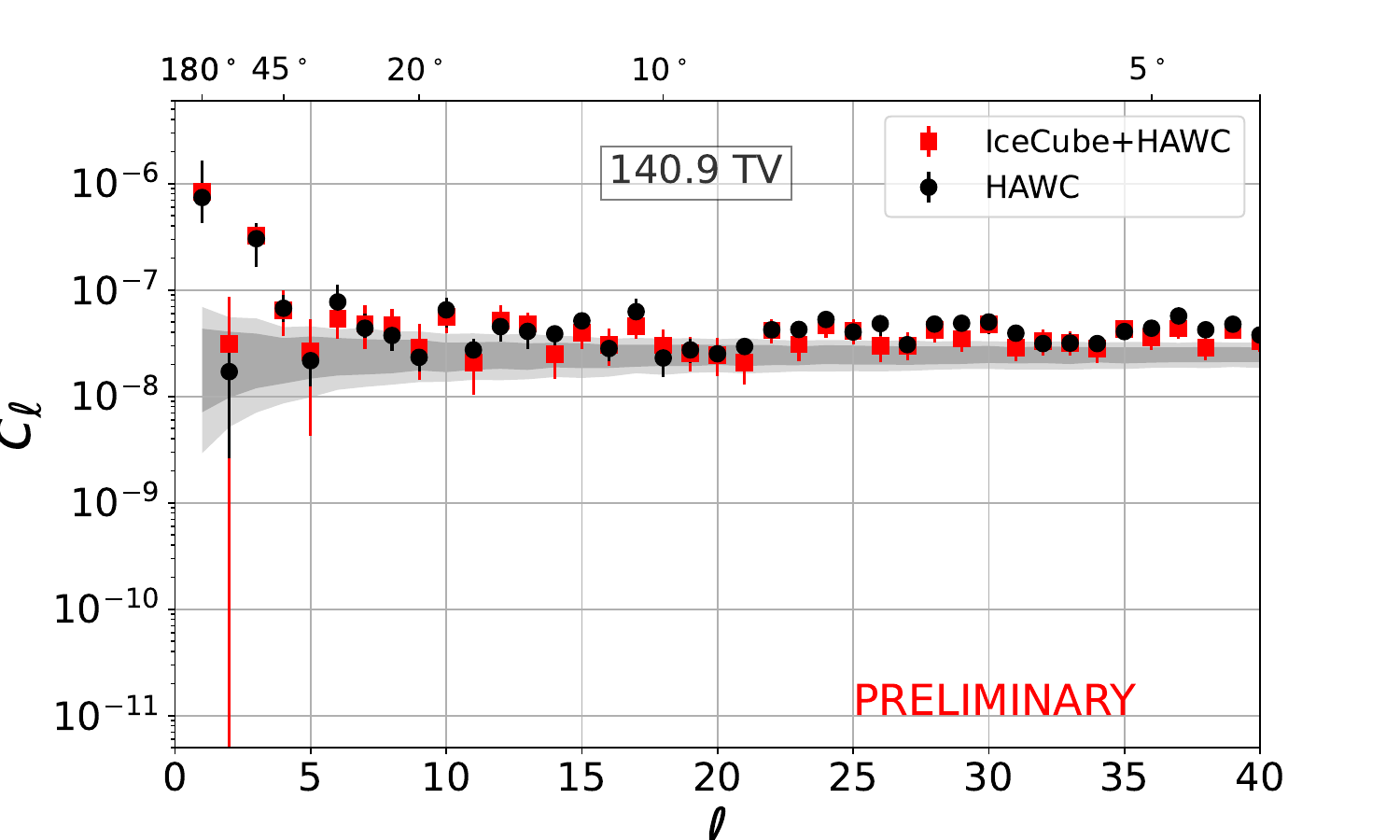}
  \includegraphics[width=0.31\textwidth]{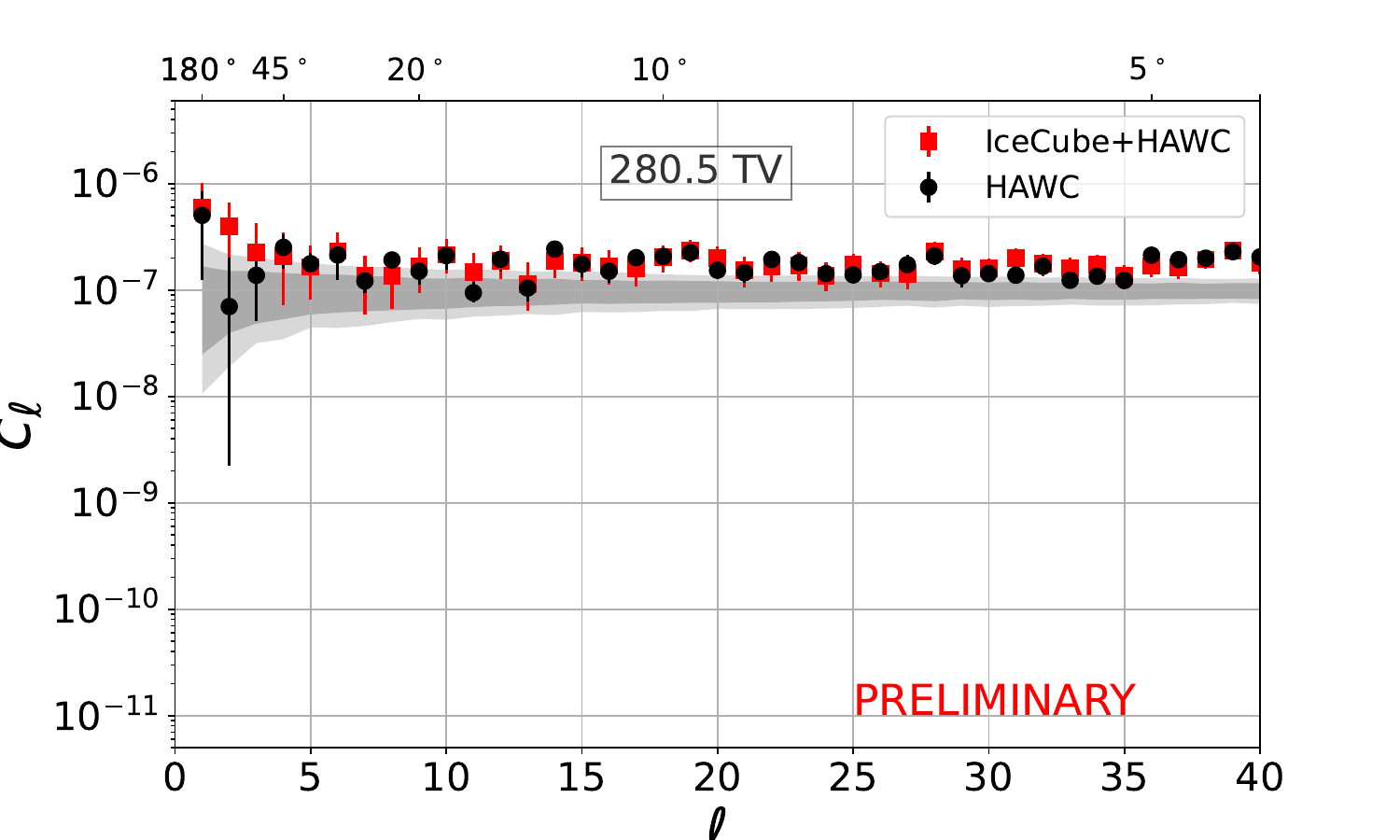}
  \caption{\footnotesize Angular power spectra for 11 HAWC sky maps binned in energy (black circles) and 7 combined all-sky IceCube+HAWC maps using rigidity-driven pairs of energy bins (red squares). Error bars represent statistical uncertainties. The shaded regions correspond to the isotropic noise at 68\% (dark) and 90\% (light) confidence intervals.}
  \label{fig:aps}
\end{figure*}

The angular power spectrum (APS) defined as 
\begin{equation}
  \mathcal{C}_{\ell} = \frac{1}{2 \ell + 1} \sum_{m=-\ell}^{\ell} | a_{\ell m} |^{2}~~,
   \label{eq:cldef_true}
\end{equation}
for each value of $\ell$,
provides an estimate of the significance of structures at different angular scales of $\sim$ 180$^\circ/\ell$. 
In case of full $4\pi$ sky coverage, the multipole moments ${a}_{\ell m}$ would give complete representation of the anisotropy. However, partial sky coverage results in large correlations between coefficients $a_{\ell m}$, biasing results. Fig.~\ref{fig:aps} shows the pseudo-angular power spectrum for all 11 HAWC rigidity maps. For maps above 4.1 TV, where we have IceCube--HAWC combined data, we include the combined angular power spectrum shown in red. The dipole power of the combined map is consistently larger than in the HAWC-only measurement, as some power is redistributed to higher $C_\ell$ terms due to correlations in the $a_{\ell m}$ coefficients. 
The isotropic noise level is driven by the region with the lowest statistics. In general, the APS has most power distributed to lower $\ell$ harmonics, corresponding to larger angular structures (e.g., dipole, quadrupole, etc.) and decreases for larger $\ell$ to be drowned in noise. In the case of combined maps, the noise level shown is computed from the total combined dataset. In some cases, the noise level of the combined maps is lower due to the increased statistics in the overlapping FoV.

\section{Conclusions and Outlook}\label{conclusions}
We have presented preliminary results on an updated full-sky analysis of the cosmic-ray arrival direction distribution with data collected by the HAWC and IceCube observatories with complementary field of views jointly covering nearly $4\pi$ steradians.
With 8 years of data, HAWC can extend the energy range from previous results~\cite{Abeysekara_2018} to higher energies. We include updated measurements by HAWC in the energy range between 3.0 TeV and 0.5 PeV, confirming previous results of an energy-dependent anisotropy in the arrival direction distribution of cosmic rays seen by other experiments.
This analysis -- which includes recently published results from IceCube with 12 years of data -- is the first all-sky anisotropy study with primary cosmic-ray energies above 10 TeV.
The combined maps and their corresponding angular power spectra largely eliminate biases that result from partial sky coverage. 
IceCube angular structures appear to evolve faster as a function of energy than they do for HAWC maps. Monte Carlo studies suggest that the angular distributions depend on rigidity rather than energy. The best-matching energy bins are consistent with this hypothesis.
Further investigation is needed, including systematic studies that account for uncertainties in cosmic-ray composition models.
We also need to extend this study by including data from other cosmic-ray observatories in both hemispheres that extend to higher and lower energies. Some of this work is already underway through various other collaborations.

% Bibtex references:
\bibliographystyle{ICRC}
\bibliography{references}

\clearpage

\section*{Full Author List: \ HAWC Collaboration}
\scriptsize
\noindent
%first.author$^1$, 
%second.author$^2$, 
%third.author$^3$ % .... more names
%and 
%last.author$^{n}$ \\
%
%\noindent
%$^1$first.affiliation.
%$^2$second.affiliation. % .... more affiliation
%$^{m}$last.affiliation.
%\vskip2cm
\noindent

R. Alfaro$^{1}$,
C. Alvarez$^{2}$,
A. Andrés$^{3}$,
E. Anita-Rangel$^{3}$,
M. Araya$^{4}$,
J.C. Arteaga-Velázquez$^{5}$,
D. Avila Rojas$^{3}$,
H.A. Ayala Solares$^{6}$,
R. Babu$^{7}$,
P. Bangale$^{8}$,
E. Belmont-Moreno$^{1}$,
A. Bernal$^{3}$,
K.S. Caballero-Mora$^{2}$,
T. Capistrán$^{9}$,
A. Carramiñana$^{10}$,
F. Carreón$^{3}$,
S. Casanova$^{11}$,
S. Coutiño de León$^{12}$,
E. De la Fuente$^{13}$,
D. Depaoli$^{14}$,
P. Desiati$^{12}$,
N. Di Lalla$^{15}$,
R. Diaz Hernandez$^{10}$,
B.L. Dingus$^{16}$,
M.A. DuVernois$^{12}$,
J.C. Díaz-Vélez$^{12}$,
K. Engel$^{17}$,
T. Ergin$^{7}$,
C. Espinoza$^{1}$,
K. Fang$^{12}$,
N. Fraija$^{3}$,
S. Fraija$^{3}$,
J.A. García-González$^{18}$,
F. Garfias$^{3}$,
N. Ghosh$^{19}$,
A. Gonzalez Muñoz$^{1}$,
M.M. González$^{3}$,
J.A. Goodman$^{17}$,
S. Groetsch$^{19}$,
J. Gyeong$^{20}$,
J.P. Harding$^{16}$,
S. Hernández-Cadena$^{21}$,
I. Herzog$^{7}$,
D. Huang$^{17}$,
P. Hüntemeyer$^{19}$,
A. Iriarte$^{3}$,
S. Kaufmann$^{22}$,
D. Kieda$^{23}$,
K. Leavitt$^{19}$,
H. León Vargas$^{1}$,
J.T. Linnemann$^{7}$,
A.L. Longinotti$^{3}$,
G. Luis-Raya$^{22}$,
K. Malone$^{16}$,
O. Martinez$^{24}$,
J. Martínez-Castro$^{25}$,
H. Martínez-Huerta$^{30}$,
J.A. Matthews$^{26}$,
P. Miranda-Romagnoli$^{27}$,
P.E. Mirón-Enriquez$^{3}$,
J.A. Montes$^{3}$,
J.A. Morales-Soto$^{5}$,
M. Mostafá$^{8}$,
M. Najafi$^{19}$,
L. Nellen$^{28}$,
M.U. Nisa$^{7}$,
N. Omodei$^{15}$,
E. Ponce$^{24}$,
Y. Pérez Araujo$^{1}$,
E.G. Pérez-Pérez$^{22}$,
Q. Remy$^{14}$,
C.D. Rho$^{20}$,
D. Rosa-González$^{10}$,
M. Roth$^{16}$,
H. Salazar$^{24}$,
D. Salazar-Gallegos$^{7}$,
A. Sandoval$^{1}$,
M. Schneider$^{1}$,
G. Schwefer$^{14}$,
J. Serna-Franco$^{1}$,
A.J. Smith$^{17}$
Y. Son$^{29}$,
R.W. Springer$^{23}$,
O. Tibolla$^{22}$,
K. Tollefson$^{7}$,
I. Torres$^{10}$,
R. Torres-Escobedo$^{21}$,
R. Turner$^{19}$,
E. Varela$^{24}$,
L. Villaseñor$^{24}$,
X. Wang$^{19}$,
Z. Wang$^{17}$,
I.J. Watson$^{29}$,
H. Wu$^{12}$,
S. Yu$^{6}$,
S. Yun-Cárcamo$^{17}$,
H. Zhou$^{21}$,

\vskip2cm
\noindent

$^{1}$Instituto de F\'{i}sica, Universidad Nacional Autónoma de México, Ciudad de Mexico, Mexico,
$^{2}$Universidad Autónoma de Chiapas, Tuxtla Gutiérrez, Chiapas, México,
$^{3}$Instituto de Astronom\'{i}a, Universidad Nacional Autónoma de México, Ciudad de Mexico, Mexico,
$^{4}$Universidad de Costa Rica, San José 2060, Costa Rica,
$^{5}$Universidad Michoacana de San Nicolás de Hidalgo, Morelia, Mexico,
$^{6}$Department of Physics, Pennsylvania State University, University Park, PA, USA,
$^{7}$Department of Physics and Astronomy, Michigan State University, East Lansing, MI, USA,
$^{8}$Temple University, Department of Physics, 1925 N. 12th Street, Philadelphia, PA 19122, USA,
$^{9}$Universita degli Studi di Torino, I-10125 Torino, Italy,
$^{10}$Instituto Nacional de Astrof\'{i}sica, Óptica y Electrónica, Puebla, Mexico,
$^{11}$Institute of Nuclear Physics Polish Academy of Sciences, PL-31342 11, Krakow, Poland,
$^{12}$Dept. of Physics and Wisconsin IceCube Particle Astrophysics Center, University of Wisconsin{\textemdash}Madison, Madison, WI, USA,
$^{13}$Departamento de F\'{i}sica, Centro Universitario de Ciencias Exactase Ingenierias, Universidad de Guadalajara, Guadalajara, Mexico, 
$^{14}$Max-Planck Institute for Nuclear Physics, 69117 Heidelberg, Germany,
$^{15}$Department of Physics, Stanford University: Stanford, CA 94305–4060, USA,
$^{16}$Los Alamos National Laboratory, Los Alamos, NM, USA,
$^{17}$Department of Physics, University of Maryland, College Park, MD, USA,
$^{18}$Tecnologico de Monterrey, Escuela de Ingenier\'{i}a y Ciencias, Ave. Eugenio Garza Sada 2501, Monterrey, N.L., Mexico, 64849,
$^{19}$Department of Physics, Michigan Technological University, Houghton, MI, USA,
$^{20}$Department of Physics, Sungkyunkwan University, Suwon 16419, South Korea,
$^{21}$Tsung-Dao Lee Institute \& School of Physics and Astronomy, Shanghai Jiao Tong University, 800 Dongchuan Rd, Shanghai, SH 200240, China,
$^{22}$Universidad Politecnica de Pachuca, Pachuca, Hgo, Mexico,
$^{23}$Department of Physics and Astronomy, University of Utah, Salt Lake City, UT, USA, 
$^{24}$Facultad de Ciencias F\'{i}sico Matemáticas, Benemérita Universidad Autónoma de Puebla, Puebla, Mexico, 
$^{25}$Centro de Investigaci\'on en Computaci\'on, Instituto Polit\'ecnico Nacional, M\'exico City, M\'exico,
$^{26}$Dept of Physics and Astronomy, University of New Mexico, Albuquerque, NM, USA,
$^{27}$Universidad Autónoma del Estado de Hidalgo, Pachuca, Mexico,
$^{28}$Instituto de Ciencias Nucleares, Universidad Nacional Autónoma de Mexico, Ciudad de Mexico, Mexico, 
$^{29}$University of Seoul, Seoul, Rep. of Korea,
$^{30}$Departamento de Física y Matemáticas, Universidad de Monterrey, Av.~Morones Prieto 4500, 66238, San Pedro Garza García NL, México

\subsection*{Acknowledgments}
We acknowledge the support from: the US National Science Foundation (NSF); the US Department of Energy Office of High-Energy Physics; the Laboratory Directed Research and Development (LDRD) program of Los Alamos National Laboratory; Consejo Nacional de Ciencia y Tecnolog\'{i}a (CONACyT), M\'{e}xico, grants LNC-2023-117, 271051, 232656, 260378, 179588, 254964, 258865, 243290, 132197, A1-S-46288, A1-S-22784, CF-2023-I-645, CBF2023-2024-1630, c\'{a}tedras 873, 1563, 341, 323, Red HAWC, M\'{e}xico; DGAPA-UNAM grants IG101323, IN111716-3, IN111419, IA102019, IN106521, IN114924, IN110521 , IN102223; VIEP-BUAP; PIFI 2012, 2013, PROFOCIE 2014, 2015; the University of Wisconsin Alumni Research Foundation; the Institute of Geophysics, Planetary Physics, and Signatures at Los Alamos National Laboratory; Polish Science Centre grant, 2024/53/B/ST9/02671; Coordinaci\'{o}n de la Investigaci\'{o}n Cient\'{i}fica de la Universidad Michoacana; Royal Society - Newton Advanced Fellowship 180385; Gobierno de España and European Union-NextGenerationEU, grant CNS2023- 144099; The Program Management Unit for Human Resources \& Institutional Development, Research and Innovation, NXPO (grant number B16F630069); Coordinaci\'{o}n General Acad\'{e}mica e Innovaci\'{o}n (CGAI-UdeG), PRODEP-SEP UDG-CA-499; Institute of Cosmic Ray Research (ICRR), University of Tokyo. H.F. acknowledges support by NASA under award number 80GSFC21M0002. C.R. acknowledges support from National Research Foundation of Korea (RS-2023-00280210). We also acknowledge the significant contributions over many years of Stefan Westerhoff, Gaurang Yodh and Arnulfo Zepeda Dom\'inguez, all deceased members of the HAWC collaboration. Thanks to Scott Delay, Luciano D\'{i}az and Eduardo Murrieta for technical support.

\clearpage
\section*{Full Author List: IceCube Collaboration}

\scriptsize
\noindent
R. Abbasi$^{16}$,
M. Ackermann$^{63}$,
J. Adams$^{17}$,
S. K. Agarwalla$^{39,\: {\rm a}}$,
J. A. Aguilar$^{10}$,
M. Ahlers$^{21}$,
J.M. Alameddine$^{22}$,
S. Ali$^{35}$,
N. M. Amin$^{43}$,
K. Andeen$^{41}$
% ...
\\
\\
% ...
$^{10}$ Universit{\'e} Libre de Bruxelles, Science Faculty CP230, B-1050 Brussels, Belgium \\
% ...
$^{16}$ Department of Physics, Loyola University Chicago, Chicago, IL 60660, USA \\
$^{17}$ Dept. of Physics and Astronomy, University of Canterbury, Private Bag 4800, Christchurch, New Zealand \\
% ...
$^{21}$ Niels Bohr Institute, University of Copenhagen, DK-2100 Copenhagen, Denmark \\
$^{22}$ Dept. of Physics, TU Dortmund University, D-44221 Dortmund, Germany \\
% ...
$^{35}$ Dept. of Physics and Astronomy, University of Kansas, Lawrence, KS 66045, USA \\
% ...
$^{39}$ Dept. of Physics and Wisconsin IceCube Particle Astrophysics Center, University of Wisconsin{\textemdash}Madison, Madison, WI 53706, USA \\
% ...
$^{41}$ Department of Physics, Marquette University, Milwaukee, WI 53201, USA \\
% ...
$^{43}$ Bartol Research Institute and Dept. of Physics and Astronomy, University of Delaware, Newark, DE 19716, USA \\
% ...
$^{63}$ Deutsches Elektronen-Synchrotron DESY, Platanenallee 6, D-15738 Zeuthen, Germany \\
$^{\rm a}$ also at Institute of Physics, Sachivalaya Marg, Sainik School Post, Bhubaneswar 751005, India \\
% ...

\subsection*{Acknowledgments}

\noindent
The authors gratefully acknowledge the support from the following agencies and institutions:
USA {\textendash} U.S. National Science Foundation-Office of Polar Programs,
U.S. National Science Foundation-Physics Division,
U.S. National Science Foundation-EPSCoR,
U.S. National Science Foundation-Office of Advanced Cyberinfrastructure,
Wisconsin Alumni Research Foundation,
Center for High Throughput Computing (CHTC) at the University of Wisconsin{\textendash}Madison,
Open Science Grid (OSG),
Partnership to Advance Throughput Computing (PATh),
Advanced Cyberinfrastructure Coordination Ecosystem: Services {\&} Support (ACCESS),
Frontera and Ranch computing project at the Texas Advanced Computing Center,
U.S. Department of Energy-National Energy Research Scientific Computing Center,
Particle astrophysics research computing center at the University of Maryland,
Institute for Cyber-Enabled Research at Michigan State University,
Astroparticle physics computational facility at Marquette University,
NVIDIA Corporation,
and Google Cloud Platform;
Belgium {\textendash} Funds for Scientific Research (FRS-FNRS and FWO),
FWO Odysseus and Big Science programmes,
and Belgian Federal Science Policy Office (Belspo);
Germany {\textendash} Bundesministerium f{\"u}r Forschung, Technologie und Raumfahrt (BMFTR),
Deutsche Forschungsgemeinschaft (DFG),
Helmholtz Alliance for Astroparticle Physics (HAP),
Initiative and Networking Fund of the Helmholtz Association,
Deutsches Elektronen Synchrotron (DESY),
and High Performance Computing cluster of the RWTH Aachen;
Sweden {\textendash} Swedish Research Council,
Swedish Polar Research Secretariat,
Swedish National Infrastructure for Computing (SNIC),
and Knut and Alice Wallenberg Foundation;
European Union {\textendash} EGI Advanced Computing for research;
Australia {\textendash} Australian Research Council;
Canada {\textendash} Natural Sciences and Engineering Research Council of Canada,
Calcul Qu{\'e}bec, Compute Ontario, Canada Foundation for Innovation, WestGrid, and Digital Research Alliance of Canada;
Denmark {\textendash} Villum Fonden, Carlsberg Foundation, and European Commission;
New Zealand {\textendash} Marsden Fund;
Japan {\textendash} Japan Society for Promotion of Science (JSPS)
and Institute for Global Prominent Research (IGPR) of Chiba University;
Korea {\textendash} National Research Foundation of Korea (NRF);
Switzerland {\textendash} Swiss National Science Foundation (SNSF).
\end{document}